\documentclass[12pt]{report}

\usepackage{appendix}
\usepackage[a4paper,top=3cm,bottom=3cm,left=3.5cm,righ t=3cm,marginparwidth=1.75cm,headheight=22pt]{geometry}
\usepackage{amsmath}
\usepackage{cite}
\usepackage{courier}

\usepackage[export]{adjustbox}
\usepackage[nottoc,notlot,notlof]{tocbibind}
\usepackage[labelfont=bf, textfont=bf]{caption}
\usepackage{graphicx}
\usepackage{booktabs}
\usepackage[bottom]{footmisc}
\usepackage{hyperref}
\usepackage{float}
\usepackage{setspace}
\usepackage{subfigure}
\usepackage{setspace}
\usepackage{lipsum}
\usepackage{fancyhdr} 
\usepackage{url}
\usepackage{tabularx}
\usepackage[utf8]{inputenc}
\usepackage{mathptmx} 

\fancypagestyle{plain}{
\fancyhf{} 
\fancyhead[R]{\bf \small \textsl{\nouppercase{\leftmark}} \vspace{0.1in}}
\fancyfoot[R]{\thepage}

}

\begin{document}
\fontdimen2\font=0.5em
\begin{titlepage}

\begin{figure}[h!]
\centering
\includegraphics[width=1\textwidth]{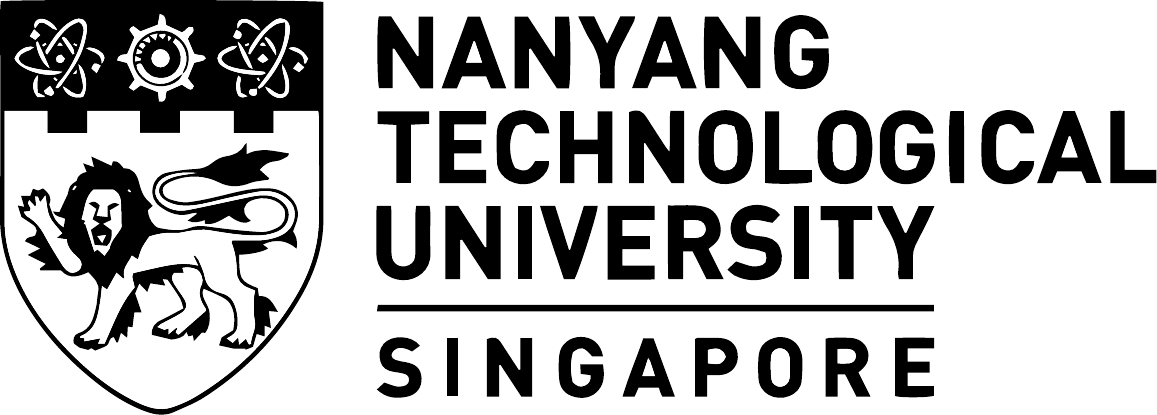}
\caption*{}
\label{fig:entropy} 
\end{figure}

\vspace{1.5in}

\centering
\Huge{\textbf{Investigation and Evaluation of Adaptive Algorithms for Multichannel Active Noise Control System}}\\[1.9in]

\LARGE{\textbf{ZHANG RUNSHENG}}\\[0.5in]

\normalsize{\textbf{SCHOOL OF ELECTRICAL AND ELECTRONIC ENGINEERING}}\\[0.2in]


\large{\textbf{2023}}
\end{titlepage}
\newpage 
\begin{titlepage}
\begin{center}
\vspace*{2in}
\Huge{\textbf{Investigation and Evaluation of Adaptive Algorithms for Multichannel Active Noise Control System}}\\[1.5in]

\LARGE{\textbf{\MakeUppercase{ZHANG RUNSHENG}}}\\[1in]

\normalsize{\textbf{\MakeUppercase{SCHOOL OF ELECTRICAL AND ELECTRONIC ENGINEERING}}}\\[0.5in]
\normalsize{\textbf{\MakeUppercase{A DISSERTATION SUBMITTED IN PARTIAL FULFILMENT OF\\THE REQUIREMENTS FOR THE DEGREE OF\\MASTER OF SCIENCE IN COMPUTER CONTROL \& AUTOMATION}}}\\[0.75in]


\large{\textbf{2023}}
\end{center}
\end{titlepage}
\newpage 
\begin{titlepage}

\begin{center}
\Large{\bf{Statement of Originality}}
\end{center}

\vspace{0.2in}

\begin{spacing}{2}

I hereby certify that the work embodied in this thesis is the result of original research, is free of plagiarised materials, and has not been submitted for a higher degree to any other University or Institution.

\end{spacing}

\vspace{2.5cm}

\begin{center}
	\makebox[4cm]{\dotfill}  \hfill \makebox[4cm]{\dotfill}\\
	\makebox[4cm]{Date}      \hfill \makebox[4cm]{Zhang Runsheng}
\end{center}
\end{titlepage}
\newpage 
\begin{titlepage}

\begin{center}
\Large{\bf{Supervisor Declaration Statement}}
\end{center}

\vspace{0.2in}

\begin{spacing}{2}
I have reviewed the content and presentation style of this thesis and declare it is free of plagiarism and of sufficient grammatical clarity to be examined.  To the best of my knowledge, the research and writing are those of the candidate except as acknowledged in the Author Attribution Statement. I confirm that the investigations were conducted in accord with the ethics policies and integrity standards of Nanyang Technological University and that the research data are presented honestly and without prejudice.
\end{spacing}

\vspace{2.5cm}

\begin{center}
	\makebox[4cm]{\dotfill}  \hfill \makebox[4cm]{\dotfill}\\
	\makebox[4cm]{Date}      \hfill \makebox[4cm]{Prof. Gan Woon Seng}
\end{center}
\end{titlepage}
\newpage 
\begin{titlepage}

\begin{center}
\Large{\bf{Authorship Attribution Statement}}
\end{center}

\vspace{0.2in}

\begin{spacing}{2}

This thesis does not contain any materials from papers published in peer-reviewed journals or from papers accepted at conferences in which I am listed as an author.

\end{spacing}

\vspace{2.5cm}

\begin{center}
	\makebox[4cm]{\dotfill}  \hfill \makebox[4cm]{\dotfill}\\
	\makebox[4cm]{Date}      \hfill \makebox[4cm]{Zhang Runsheng}
\end{center}
\end{titlepage}
\newpage 


\pagenumbering{roman}

\renewcommand*\contentsname{\centering Table of Contents}
\tableofcontents
\newpage

\newpage

\chapter*{\centering Abstract}
\markboth{Abstract}{}

\begin{spacing}{1.5}
\setlength{\parskip}{0.3in}

\addcontentsline{toc}{chapter}{Abstract}
This dissertation focuses on the investigation and evaluation of adptive algorithms for multichannel active noise control system. The aim of the research is to investigate the effectiveness of the FxLMS algorithm and the pre-trained control filter in attenuating various types of noise.

The study begins with a comprehensive review of the existing literature on active noise control, highlighting the significance of noise reduction in different applications. The theoretical foundations of the FxLMS algorithm and the pre-trained control filter are then presented, including their underlying principles and mathematical formulations. Through a comprehensive analysis, a clear understanding of these methods is established.

To assess the performance of the FxLMS algorithm and the pre-trained control filter, extensive simulation experiments are conducted using real-world noise signals. The experiments include scenarios such as aircraft noise, traffic noise, and mixed noise. The results of the simulations are used to compare the noise reduction capabilities of the two methods and to evaluate their effectiveness under different noise conditions.

The findings indicate that the FxLMS algorithm exhibits remarkable noise reduction performance. It demonstrates a strong ability to track and respond quickly to varying noise patterns. In the initial stages of noise reduction, the pre-trained control filter shows better performance. However, as time progresses, the FxLMS algorithm consistently achieves higher average noise reduction levels compared to the pre-trained control filter. Additionally, the FxLMS algorithm shows faster responsiveness during transitional periods when noise characteristics change. These findings contribute to the field of noise control and provide valuable insights for designing efficient noise reduction systems.

\end{spacing}
\newpage


\chapter*{\centering Acknowledgements}
\markboth{Acknowledgements}{}
\begin{spacing}{1.5}
\setlength{\parskip}{0.3in}
\addcontentsline{toc}{chapter}{Acknowledgement}

I would like to thank my supervisor, Professor Gan Woon Seng, for his guidance, support, and valuable insights throughout the entire research process. his expertise and encouragement have been instrumental in shaping the direction of this study.

I would like to express my heartfelt gratitude to my senior colleagues, Professor Shi Dongyuan and PhD Luo Zhengding, for their invaluable guidance and support throughout my academic journey.

Lastly, I would like to express my heartfelt thanks to my friends and family for their understanding, encouragement, and unwavering support throughout this research journey.
\end{spacing}
\newpage

\chapter*{\centering Acronyms}
\markboth{Acronyms}{}
\begin{spacing}{1.5}
\setlength{\parskip}{0.3in}
\addcontentsline{toc}{chapter}{Acronyms}

\begin{table}[ht]
\centering
\begin{tabular}{ll}
\textbf{ANC} & Active Noise Control \\
\textbf{McANC} & Multichannel Active Noise Control \\
\textbf{FIR} & Finite Impulse Response \\
\textbf{DSP} & Digital Signal Processing \\
\textbf{JMC} & Joint Monitoring and Control \\
\textbf{FxLMS} & Filtered-x Least Mean Square \\
\textbf{NN} & Neural Network \\
\textbf{ANN} & Artificial Neural Networks\\
\textbf{VFxLMS} & Volterra Filtered-x Least Mean Square\\
\textbf{FsLMS} & Filtered-s Least Mean Square\\
\textbf{FxAP} & Filtered-x Affine Projection\\
\textbf{MSE} & Mean Square Error\\
\textbf{FxLMP} & Filtered-x Least Mean P-norm\\
\textbf{FxlogLMS} & Filtered-x Logarithmic Transformation Least Mean Square\\
\textbf{FLANN} & Functional Link Artificial Neural Network\\
\textbf{RFsLMS)} & Robust Filtered-s Least Mean Square\\
\textbf{SPL} & Sound Pressure Level\\
\textbf{IIR} & Infinite Impulse Response\\
\textbf{LMS} & Least Mean Square\\
\textbf{MMSE} & Minimum Mean Square Error\\
\textbf{MAC} & Multiplication and Accumulation\\
\textbf{RAM} & Random Access Memory\\
\textbf{SNR} & Signal-to-noise Ratio\\
\textbf{} & \\
\end{tabular}%
\end{table}

\end{spacing}
\newpage

\chapter*{\centering Symbols}
\markboth{Symbols}{}
\begin{spacing}{1.5}
\setlength{\parskip}{0.3in}
\addcontentsline{toc}{chapter}{Symbols}

\begin{table}[ht]
\centering
\begin{tabular}{ll}
\textbf{${p_1}$} & the instantaneous sound pressure value\\
\textbf{${E_1}$} & the average incident sound energy density\\
\textbf{$\Delta$} & the difference in SPL\\
\textbf{$E\left[  \cdot  \right]$} & the expectation operation\\
\textbf{${\left[  \cdot  \right]^T}$} & the transposition of the matrix\\
\textbf{$\sum$} & the summation\\
\textbf{$\approx $} & the approximate equality sign\\
\textbf{${\left[  \cdot  \right]^{ - 1}}$} & the inverse matrix\\
\textbf{$\hat \nabla $} & the instantaneous gradient\\
\textbf{$\partial $} & the partial differential equation\\
\end{tabular}%
\end{table}

\end{spacing}
\newpage

\renewcommand{\listfigurename}{\centering List of Figures}
\listoffigures
\addcontentsline{toc}{chapter}{Lists of Figures}
\newpage

\renewcommand{\listtablename}{\centering List of Tables}
\listoftables 
\addcontentsline{toc}{chapter}{Lists of Tables}
\newpage



\pagenumbering{arabic}

\chapter{Introduction}
\begin{spacing}{1.5}
\setlength{\parskip}{0.3in}
\section{Background}
In the physical sense, noise refers to sound waves that are completely irregular, with no discernible pattern in either amplitude or frequency. From an environmental perspective, however, noise is broadly defined as any sound that is considered undesirable by humans. As society and the economy have progressed, people have begun to demand a higher quality of life, including environmental standards. Noise from various aspects of modern life, such as traffic and industrial production, is now a significant disturbance to people's lives.

Research has shown that prolonged exposure to high-intensity noise can lead to increased fatigue\cite{stansfeld2003noise}, low mood and a generally negative attitude. It can also cause varying degrees of damage to various human organs, including hearing loss, digestive disorders, endocrine disruption, high blood pressure and cardiovascular disease. Noise pollution has been identified by the United Nations as one of the four major types of pollution affecting humanity, along with water pollution, air pollution and solid waste pollution\cite{basner2014auditory}. Therefore, noise has become an urgent issue that we need to address.

Traditional noise control techniques primarily focus on acoustical methods for noise control. The main methods include absorption, insulation, and the use of mufflers\cite{elliott2000signal}. These noise control methods work on the principle of dissipating sound energy through the interaction between sound waves and acoustical materials or structures, aiming to achieve noise reduction. 

These passive control methods are effective in controlling medium to high frequency noise. However, when it comes to controlling low-frequency noise, passive control devices are typically bulky, heavy, and have limited applicability. To address low-frequency noise control, active noise control (ANC) technology offers a promising solution~\cite{lam2021ten,shi2023active}.

ANC techniques utilize electronic systems to actively generate anti-noise signals that cancel out the undesired noise. By sensing the incoming noise and producing an appropriate anti-noise signals, ANC systems can effectively attenuate low-frequency noise\cite{elliott2000signal,shi2020feedforward,jiang2021modified}. 

\section{Introduction of Active Noise Control}
The mechanism of ANC involves the artificial  generation of a secondary sound signals within a defined area to control the primary sound signals\cite{widrow1982signal}. Based on the conditions of destructive interference between two sound waves, if an artificially added secondary sound source produces sound waves of equal amplitude and opposite phase to the primary sound source, the two sound waves will spatially combine coherently to form a quiet zone where the noise is cancelled out. 

In general, the ANC systems can be classified into three main types: feedforward structure, feedback structure, hybrid structure~\cite{shen2021alternative}, and multichannel structure\cite{shi2020feedforward}. In a feedforward ANC system, a reference signal is obtained from the noise source, and a control signal is generated and applied to the secondary sound source. 

In a feedback ANC system, the control signal is obtained only by filtering the error signal. The error signal is obtained by comparing the desired signal with the output signal of the system.

The multichannel active noise control (McANC) system is developed based on the first two structural models. Digital filters are often used in digital controller to implement specific algorithms, typically adaptive algorithms\cite{hansen2002understanding}. This makes them suitable for achieving noise cancellation in multichannel and time-varying environments.

\subsection{Multichannel Active Noise Control System}

In many practical application environments, there are often multiple noise sources rather than just one, such as in aircraft or car cabins. The sound sources are diverse and widely distributed. In such cases, using only one input microphone, one loudspeaker and one error microphone for adjustment would not achieve the desired noise reduction effect. Therefore, researchers have proposed the concept of McANC\cite{elliott2000signal, tanaka2014multi,shi2017understanding, luo2022implementation}. Due to the complexity of both structures and algorithms, more advanced ANC theory and algorithms are required to achieve the noise reduction process.

\begin{figure}[ht]
\centering
\includegraphics[width=10cm]{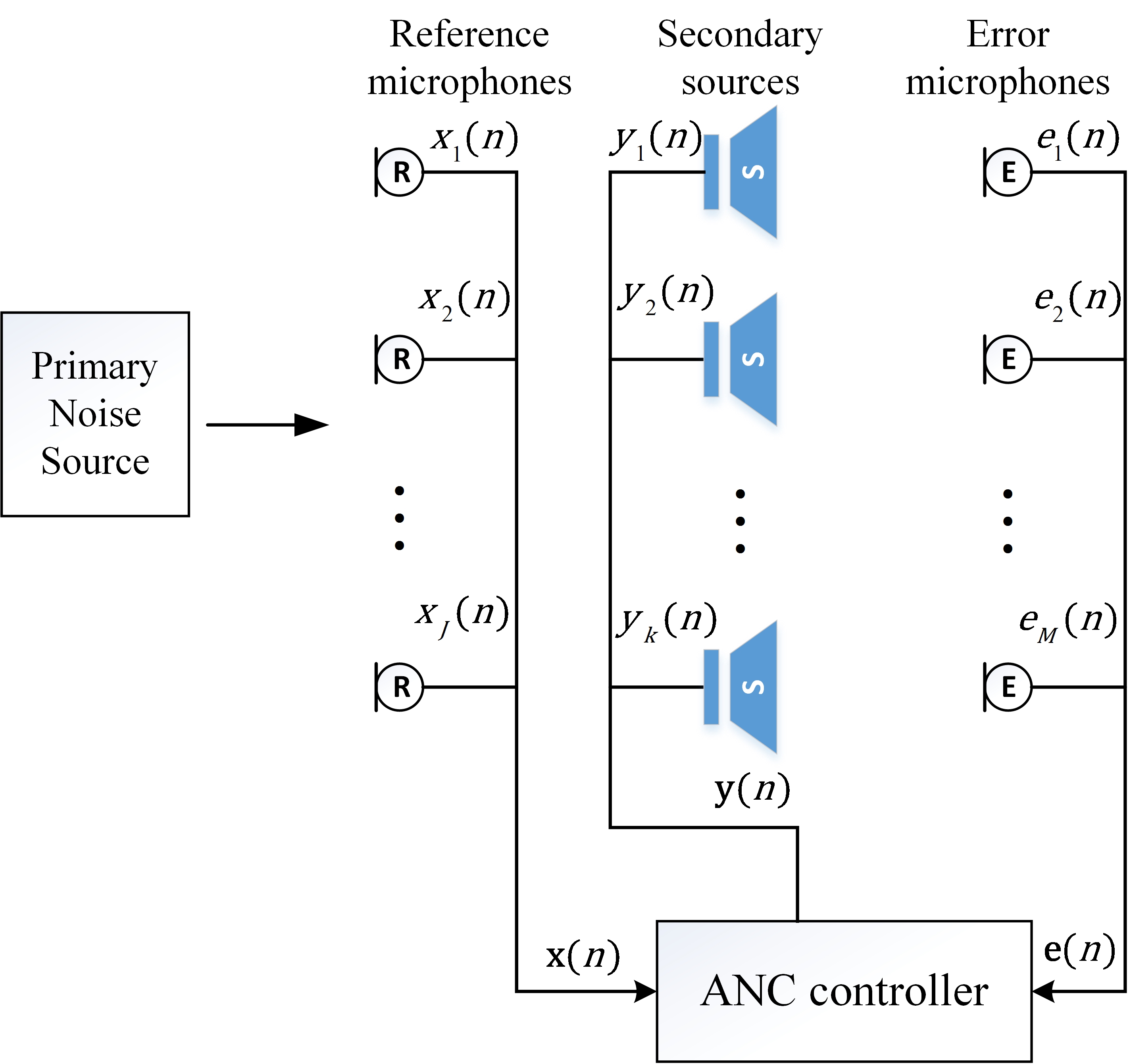}
\caption{Block diagram of the multichannel active noise control (McANC) system~\cite{shi2020algorithms}.}
\label{fig:multichannel system} 
\end{figure}

A McANC system has a number of input terminals, and the arrangement of each of the input terminals can be different. They can be arranged longitudinally along the sound channel or distributed in a planar or three-dimensional space. The block diagram of a McANC system is shown in \autoref{fig:multichannel system}. In general, the input and output of a McANC system consists of an array of sensors. 

In practical applications, feedforward ANC has proven to be highly effective in dealing with broadband noise\cite{shi2020algorithms,shi2023computation}. Its ability to effectively attenuate noise across a wide range of frequencies has contributed to its widespread utilization and implementation in numerous applications, such as multichannel ANC windows~\cite{shi2016open, shi2017algorithms,hasegawa2019multi,lam2018active,shi2019practical,he2019exploiting,lam2020active,lam2020active1}.

\subsection{Fixed Filter Techniques}
Fixed filter is a type of digital filter with pre-trained characteristics. This filter is designed to attenuate specific frequency components of the noise signal. 

Fixed control filter, particularly finite impulse response (FIR) filter, offers several advantages in ANC applications. First, it provides a computationally efficient solution, as the filter coefficients are determined offline and do not require continuous adaptation during real-time operation. This reduces the computational complexity of the ANC system, making it more suitable for real-time implementation on hardware platforms with limited resources.

Additionally, pre-trained FIR filter offers stability and robustness in ANC systems\cite{ranjan2016selective,shi2018novel}. Since the filter coefficients are fixed, the ANC system is not susceptible to sudden changes or disturbances in the noise characteristics. This stability ensures consistent noise reduction performance over time, even in dynamic noise environments. However, the fixed-filter ANC technique only employs the pre-trained control filter based on a particular acoustic environment; as a result, its noise reduction efficacy is sensitive to the variation of the environments. The selective fixed-filter ANC (SFANC) method has been proposed as a solution to this problem~\cite{shi2018novel, shi2020feedforward, wen2020improved,wen2020using}. It can automatically swap the pre-trained control filter based on the primary noises, which substantially improves the fixed-filter method's noise reduction performance~\cite{shi2022selective,luo2022hybrid,luo2023performance,shi2023transferable,luo2023deep}. 

\subsection{Adaptive Filter Techniques}
Given the dynamic nature of noise sources and environmental factors, along with the impact of these factors on the control system, achieving a real-time ANC system capable of effectively tracking and adapting to changes in noise sources and environmental conditions presents a significant challenge.

\begin{figure}[ht]
\centering
\includegraphics{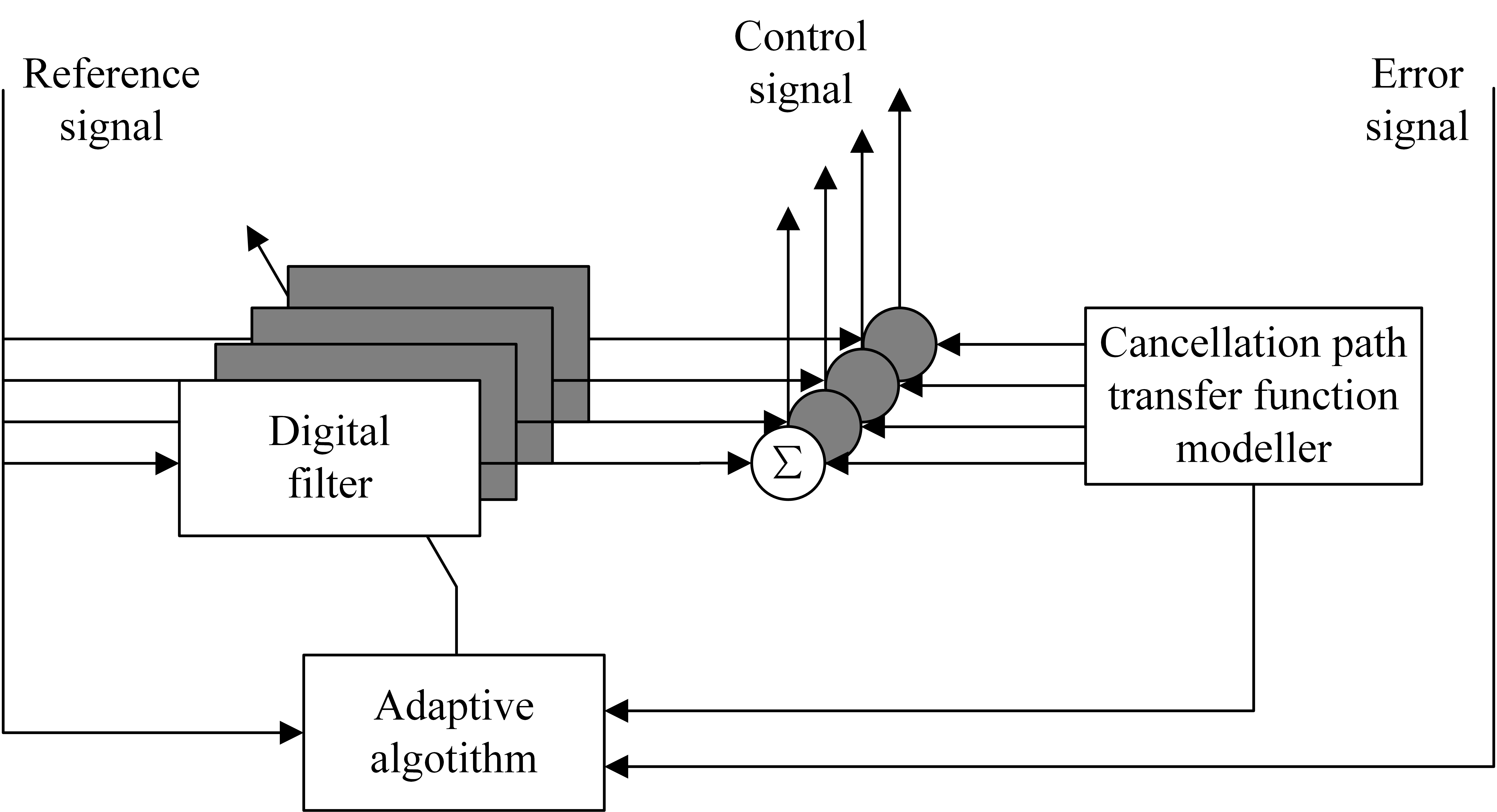}
\caption{Basic components of the adaptive filter of a feedforward McANC  system.}
\label{fig: adaptive filter} 
\end{figure}

To achieve this goal, researchers have applied the principles of adaptive control. One of the most commonly used techniques in this context is the adaptive filter. The ANC system can continuously adjust its parameters and characteristics based on the measured noise signals and the desired output\cite{snyder2000active}. This allows the system to dynamically track and respond to changes in noise sources and environmental factors, enabling real time adjustment of the secondary sound signals to effectively reduce noise. A typical adaptive filter of feedforward a McANC system is shown in \autoref{fig: adaptive filter}

During the adaptive control process, the controller first needs to capture the noise source signals using high-precision microphones. The analog noise signals are then converted to a digital signals that can be recognised by a digital signals processor via an analog to digital converter.

The digital signals processor applies adaptive filtering algorithms to produce secondary signals. This secondary signals have the same amplitude and frequency as the original noise signals, but with the opposite phase. After a series of circuit processes, the secondary signals are emitted through the output loudspeakers, where they combine with the original noise in the acoustic field. The error microphones capture the combined sound wave signals and provide real time feedback to the McANC system. This feedback is used to continuously update and iterate the adaptive control filter coefficients until an active quiet zone is created around the error microphones. This entire process is considered, including the sound wave transmission channel, as a system.

The performance of this system is influenced by various factors, including the hardware circuits, the capabilities of the digital signal processor, and notably, the effectiveness of the applied adaptive filtering algorithm.

\section{Objectives of Work}
In this dissertation, a thorough investigation of the current state-of-the-art algorithms that are being used for McANC system is performed. Initially, a comprehensive analysis of the noise reduction performance and computational complexity of adaptive algorithms will be conducted through a literature review. Subsequent simulation experiments will compare the noise reduction performance of different adaptive algorithms on real-world noise signals. Through thorough results analysis, valuable guidance directions will be provided for the design of practical noise reduction systems.

\section{Organisation of the Dissertation}
Chapter 1 provides an introduction to the research topic, including the background and basic concepts of related theory. Chapter 2 critically examines the existing body of knowledge on active noise control and discusses various noise control algorithms and techniques. Chapter 3 presents the theoretical foundations and concepts necessary to understand the research methodology. Chapter 4 presents the simulation results obtained from applying the proposed methodology. Chapter 5 summarizes the key findings of the research and future work.

\end{spacing}
\newpage


\chapter{Literature Review}
\begin{spacing}{1.5}
\setlength{\parskip}{0.3in}

In order to conduct a comprehensive investigation into the state-of-the-art algorithms employed in the McANC system and to select suitable algorithms for simulation, this chapter provides an exhaustive literature review of ANC technology and adaptive algorithms for the control filter. The current development status of ANC technology provides theoretical support for determining the structure of the McANC system, while the current development status of adaptive algorithms offers insights for algorithms selection.

\section{ANC Technology}

In the 1930s, P. Lueg, a German researcher, first introduced and patented the concept of ANC technology\cite{lueg1936process}. This early concept of noise control inspired subsequent researchers to delve deeper into the field and explore further innovative solutions.

In 1956, W. Conover proposed the use of ANC to control transformer noise. Transformer noise is characterised by prominent single frequency components and clear periodicity, making it well suited to ANC techniques\cite{conover1956fighting}.

These early developments demonstrated the effectiveness and versatility of ANC technology in addressing specific noise problems. However, for a long period in the 1960s, there were few successful studies on the practical application of ANC technology, leading to a relatively quiet phase in ANC research.

On the one hand, the understanding of the noise reduction mechanism of ANC was not deep enough at that time. On the other hand, the implementation of ANC systems relied on analog circuitry, which posed significant challenges in practical applications. The main reasons for these challenges included:

\begin{enumerate}
\item  The characteristics of real world noise are almost always time-varying.
\item Parameters in the control circuit and acoustic environment often change over time.
\item Complex noise sources and the need for multichannel systems to extend the noise reduction range result in more complex transfer functions for the controller.
\end{enumerate}

These difficulties could not be overcome with analog circuitry, so ANC research was limited to simple experimental stages with little progress during this period.

In the 1970s, with the rapid development of large-scale integrated circuits and digital signal processing (DSP) techniques, the barriers to the development of ANC technology began to be overcome. Kido in 1975  was the first to apply digital technology to ANC. The widespread use and rapid development of DSP techniques and devices made it possible to implement practical ANC systems\cite{kido1975reduction,shi2016comparison}. Since then, some high-performance processors, such as Field Programmable Gate Arrays (FPGA), have been gradually implemented to realize the more complex multichannel ANC system in order to attain a higher sampling rate and enhanced noise reduction performance~\cite{shi2016systolic,shi2017multiple}. 

Gradually, several theories and technologies have been applied to people's daily lives. In 1988, researchers at the University of Southampton applied ANC to aircraft, resulting in a 10-15 dB reduction in internal aircraft noise and significant noise reduction effects\cite{elliott2000signal,kuo2006active,elliott1987multiple,kuo1995development}. In 1995, the US company Lotus introduced ANC to automobiles, resulting in a reduction in internal vehicle noise and improved passenger comfort. This breakthrough in automotive noise reduction technology paved the way for the future application of ANC systems in transport vehicles. 

At present, ANC systems have been widely implemented in various aspects of practical life. Professor Sen M. Kuo has conducted research on improved adaptive algorithms and filter structures. He has applied these findings to various ANC systems\cite{chang2016listening}, such as infant incubators and automotive cockpits. The application goal of the ANC system in infant incubators is to create a quiet zone near the baby's ears. This requires the use of two auxiliary loudspeakers and two error microphones. One of the main challenges in this application is that the noise sources come from both inside and outside the incubator, resulting in multiple noise sources. Over the course of nearly two decades, Professor Sen M. Kuo and his team have conducted research on integrating ANC systems in various scenarios, including noise-canceling pillows, open-fit hearing aids, and noise-canceling headphones~\cite{shen2021wireless, shen2021implementation, shen2022adaptive, shen2022hybrid, shen2022multi, shen2023implementations}.

In 2015, Sakamoto developed a feedback-based ANC technology that can simultaneously reduce multiple narrowband noises, specifically for road noise\cite{sakamoto2015development}. The research focused on an adaptive filter called the multi-frequency adaptive notch filter and explored its connection method. Based on this, they proposed a method of connecting multiple filters to minimize the mutual interference caused by different controller transfer characteristics. This approach enabled the design of multiple narrowband phase-shifting controllers, ultimately achieving the desired noise reduction.

In 2020, Lam Bhan proposed a residential ANC system installed on windows that aimed to attenuate urban traffic noise while maintaining natural ventilation\cite{lam2020active}. The active control implementation in their window system resembled a multi-channel version of noise-canceling headphones. The system successfully reduced the energy-averaged sound pressure level of typical urban noise by up to 10dB.

\section{Adaptive Algorithms for the Control Filter}

Since the 1940s, the theory and technology of adaptive filtering have gradually matured, providing a breakthrough solution for the "automatic adaptation to the environment" requirement of ANC controllers and addressing the time-varying nature of ANC systems~\cite{lu2021survey,ji2023practical}. The work on adaptive ANC was first conducted by Oppenheim et al. from General Electric Company between 1957 and 1960\cite{oppenheim1994single, ingle2005statisical}. 

During their research, Oppenheim and his team introduced the concept and principles of adaptive filter and applied them to the field of noise control for the first time. This successful research marked the beginning of a new era in adaptive ANC and laid the foundation for subsequent studies and applications. Adaptive filter adjusts the parameters of the filter in real time based on the characteristics of the environmental noise signal, allowing the control system to adapt to different noise environments and achieve effective noise suppression. This technology significantly improves the performance and practicality of ANC systems, making them a crucial research direction in the field of noise control.

In the 1970s, Nelson developed the joint monitoring and control (JMC) algorithm for active noise cancellation in three-dimensional space and successfully applied it in transformer noise reduction experiments\cite{nelson1991active}. In the 1980s, Burgess et al. began research into adaptive three-dimensional ANC, leading to the proposal of the classic filtered-x least mean square (FxLMS) algorithm\cite{burgess1981active}. In the mid-1980s, Professor Elliott and his colleagues investigated the placement of secondary noise sources and monitoring microphones in ANC systems\cite{elliott1993active}. This was subsequently followed by successful flight tests of ANC in enclosed aircraft cabins.

In the late 1990s, with the advancement of neural network (NN) technology\cite{lu1994artificial,bouchard1999improved,bouchard2001new}, researchers began to incorporate it into ANC to address the adaptive noise cancellation problem caused by nonlinear cross-coupling\cite{bouchard2001new}. The application of artificial neural networks (ANN) to ANC solved the problem of system instability often encountered in traditional adaptive ANC applications and initiated a new chapter in ANC research.

In recent years, various types of nonlinear ANC algorithms have been proposed to address the issue of primary and secondary channel nonlinear distortion commonly encountered in ANC systems. These include the Volterra filtered-x least mean square (VFxLMS) algorithms\cite{tan2001adaptive,sicuranza2004filtered}, filtered-s least mean square (FsLMS) algorithms\cite{das2004active,reddy2008fast,reddy2009fast,guo2018sparse}, kernel adaptive filtering algorithms\cite{bao2009active, liu2018kernel}, wavelet algorithms\cite{qiu2016multi,akraminia2015active,akraminia2017feedforward}, spline adaptive filtering algorithms\cite{patel2015nonlinear,patel2016compensating}, neural network algorithms\cite{george2012reduced,george2013active}, and more.

In 2001, Professor Tan proposed the VFxLMS algorithm for feedforward ANC based on a multichannel structure\cite{tan2001adaptive}, which outperforms the FxLMS algorithm when dealing with nonlinear distortion in the primary channel and non-minimum phase characteristics in the secondary channel. In 2004, Professor Sicuranza and colleagues introduced the filtered-x affine projection (FxAP) algorithm based on second-order Volterra filters\cite{sicuranza2004filtered}, which can be extended to higher-order Volterra kernels for McANC. Considering the computational complexity, Das et al. designed the FsLMS algorithm and demonstrated that when the length of the secondary channel is less than 6, the computational complexity of the FsLMS algorithm is significantly lower than that of the VFxLMS algorithm\cite{das2004active}.

Furthermore, for the presence of impulsive noise in practical applications, algorithms based on the mean square error (MSE) criterion may encounter divergence issues. The filtered-x least mean p-norm (FxLMP) algorithm addresses the stability problem of ANC algorithms in impulsive noise environments\cite{leahy1995adaptive}. The filtered-x logarithmic transformation least mean square (FxlogLMS) algorithm exhibits stronger robustness against impulsive noise and does not require prior information about the noise\cite{wu2010active}. A novel robust FsLMS algorithm based on the functional link artificial neural network (FLANN) structure, called robust filtered-s least mean square (RFsLMS) algorithm, further enhances the algorithm's performance. Compared to the FxlogLMS algorithm, the RFsLMS algorithm achieves similar control effects for Gaussian and non-Gaussian noise processes\cite{george2012robust}. Furthermore, to overcome the non-linearity of the secondary path, such as output saturation~\cite{shi2017effect}, some constraint adaptive algorithms have been developed to ensure the linear operation of the ANC system~\cite{shi2019two,shi2019practicalA,shi2019optimal,shi2021optimal,shi2021comb, shi2021optimal1,shi2023frequency,shi2023multichannel,shen2023momentum,lai2023mov}.

\section{Summary}

This chapter highlighted the importance of ANC in various fields and provided an overview of existing research and advancements in the field.

Section 2.1 presented the historical evolution of ANC technology, encompassing the challenges encountered during its development and the contemporary applications of ANC technology across various domains.

Section 2.2 underscored the significance of adaptive algorithms in noise control, showcasing the endeavors of previous researchers in directions such as reducing computational complexity and enhancing stability.

\end{spacing}
\newpage


\chapter{Related Theory of ANC}
\begin{spacing}{1.5}
\setlength{\parskip}{0.3in}

Adaptive ANC systems can be divided into two types: feedforward and feedback, based on their control configuration. Feedforward systems require a reference signal and control is achieved by a feedforward filter. On the other hand, feedback systems operate without a reference signal and the entire system relies on an error sensor that simultaneously measures the reference and error signals. In general, feedforward systems are preferred whenever possible due to their superior stability compared to feedback systems. The algorithms for ANC discussed in this dissertation are implemented using digital controllers based on feedforward control.

The core components of an adaptive ANC system are the adaptive filter and the corresponding adaptive algorithms. The following sections provide a detailed explanation of the relevant theoretical techniques used in this study. 

\section{Acoustic Mechanism of Active Noise Control}\label{Acoustic Mechanism of Active Noise Control}

Sound propagation in air occurs through the compresssion and rarefaction of air particles, resulting in the vibration of the medium. The time taken for one complete vibration is known as a cycle, and the number of cycles per second is referred to as the frequency. The degree to which air is compressed and expanded by the sound wave is called the amplitude. The phase of a sound wave refers to its position within one cycle.

\begin{figure}[H]
\centering
\includegraphics{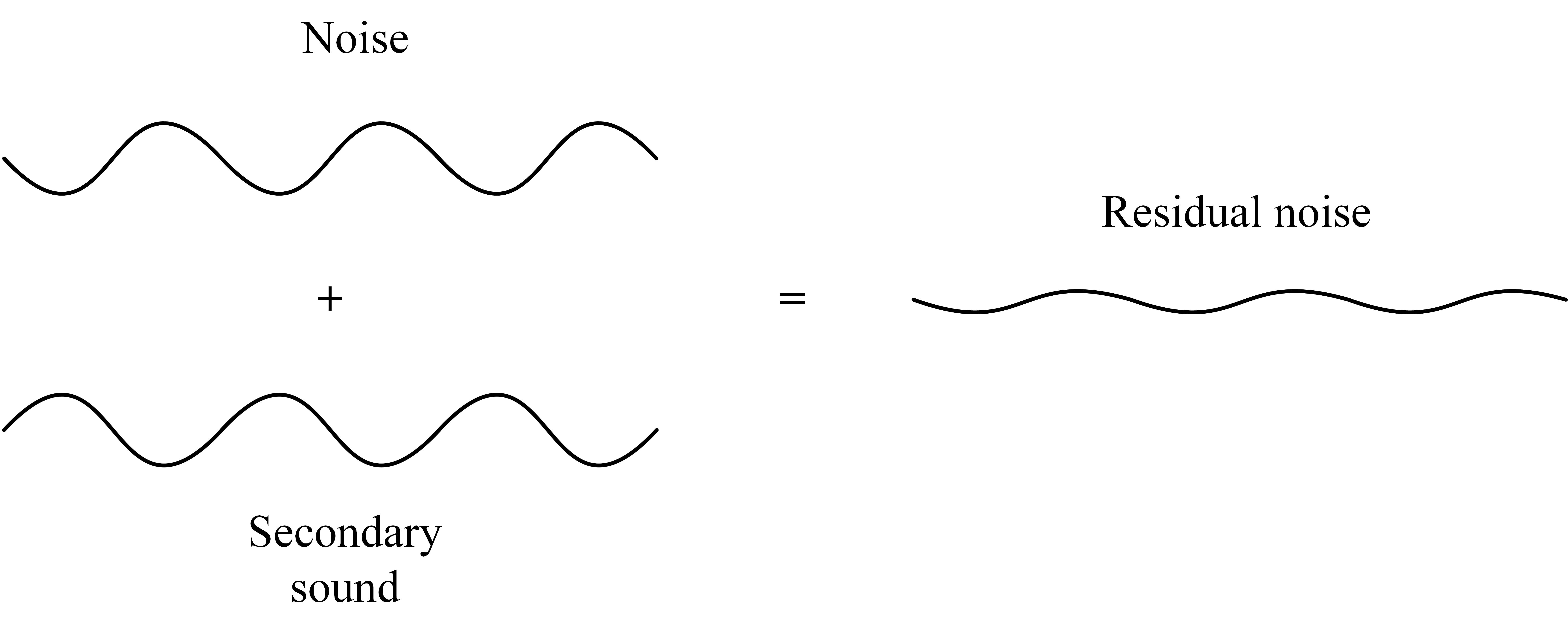}
\caption{Principle of acoustic destructive interference of sound waves.}
\label{fig:sound wave} 
\end{figure}

ANC is a technology based on the principle of wave interference, specifically the phenomenon of Yang's interference in sound waves . It involves the intentional addition of a secondary sound source within a designated area to control the primary noise signal. According to the conditions of destructive interference between two sound waves, if the amplitudes of the two signals are equal but out of phase (180 degrees apart), they will interfere with each other in space, creating a zone of reduced sound known as the "quiet zone."\cite{snyder2000active} This phenomenon is referred to as destructive interference of sound waves shown in \autoref{fig:sound wave}.

The incident sound wave (primary noise signal) can be represented as\cite{snyder2000active}:
\begin{equation}
p_1=A \cos (\omega \mathrm{t}-\varphi)
\end{equation}
where A denotes the sound pressure amplitude, $\omega$ denotes the sound wave frequency, and $\varphi$ represents the phase value. The average incident sound energy density can be expressed as:
\begin{equation}
E_1=\frac{A^2}{4 \rho c^2}
\end{equation}
where $\rho$ represents the density of the medium, and $c$ represents the speed of sound in air. Additionally, a secondary sound signal is introduced to satisfy the condition for sound wave coherence as:
\begin{equation}
{p_2} = \beta A\cos \left( {\omega \mathrm{t} - \varphi  + \alpha } \right).
\end{equation}
The combined sound energy density after the superposition of the two sound waves is given by:
\begin{equation}
{E_2} = {{{A^2}} \over {4\rho {c^2}}}\left( {1 + 2\beta \cos \alpha  + {\beta ^2}} \right)
\end{equation}
where $\alpha$ represents the phase difference between the two waves, $\beta$ denotes the amplitude ratio between the two waves. The difference in sound pressure level (SPL) between the primary sound wave and the superposed secondary sound wave at any point in the air is given by:
\begin{equation}
    \Delta  = 10\lg {{{E_1}} \over {{E_2}}} =  - 10\lg \left( {1 + 2\beta \cos \alpha  + {\beta ^2}} \right).
\end{equation}
As evident from the above equation, when the amplitudes of the two sound waves are nearly equal as $\beta$ approaches 1 and their phases are nearly opposite as $\alpha$ approaches $\pi$, the primary noise signal undergoes significant attenuation due to mutual cancellation with the secondary sound signal in the corresponding spatial region. This forms the theoretical foundation of ANC technology.

\section{Adaptive Control}\label{Adaptive Control}

The process of adaptive control consists of two main components: algorithm design and filter design. The filter is primarily used for filtering purposes, while the algorithm is responsible for adjusting the filter's transfer function or impulse response. In various application scenarios, noise is constantly changing. In order to achieve better noise reduction, this dissertation focuses on the research and implementation of adaptive algorithms.

\section{Adaptive Filter}\label{Adaptive Filter}

In signal processing, the use of filters is essential to generate the desired output. Active noise control technology requires the controller to track the variations in the primary noise signal in real-time and generate a secondary signal with the same amplitude, frequency, and a phase difference of 180 degrees. To achieve this, the controller needs to automatically adjust the system parameters based on past and current external inputs to achieve optimal performance. This type of controller is known as an adaptive filter.

\begin{figure}[ht]
\centering
\includegraphics{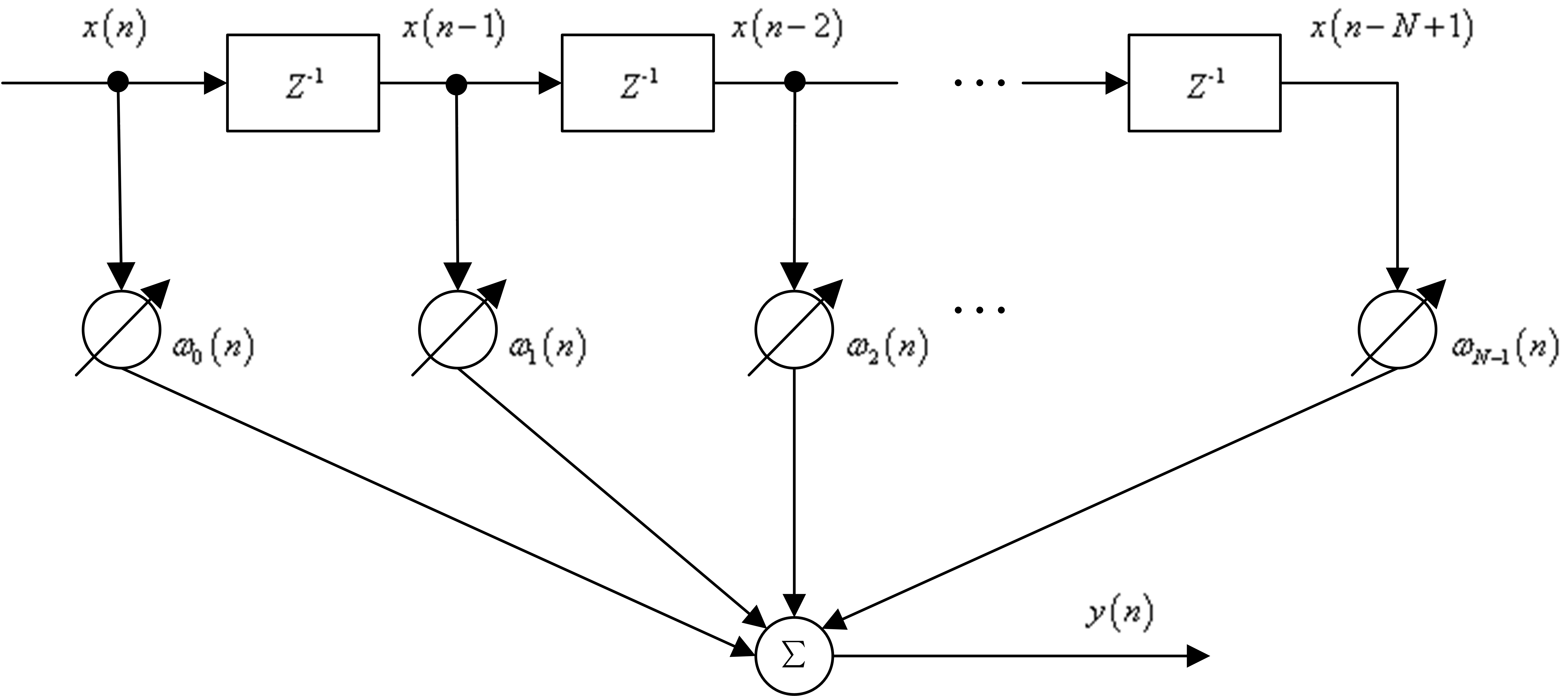}
\caption{Structure of an FIR filter.}
\label{fig:FIR} 
\end{figure}

Adaptive filters can be classified into two main types based on the superposition principle of their input and output signals. These types are linear filters and nonlinear filters. In the context of ANC systems, linear filters are commonly employed. Linear filters can be further categorized into two subtypes: FIR filters and infinite impulse response (IIR) filters. \autoref{fig:FIR} depicts the structure of an FIR filter.

\begin{figure}[ht]
\centering
\includegraphics{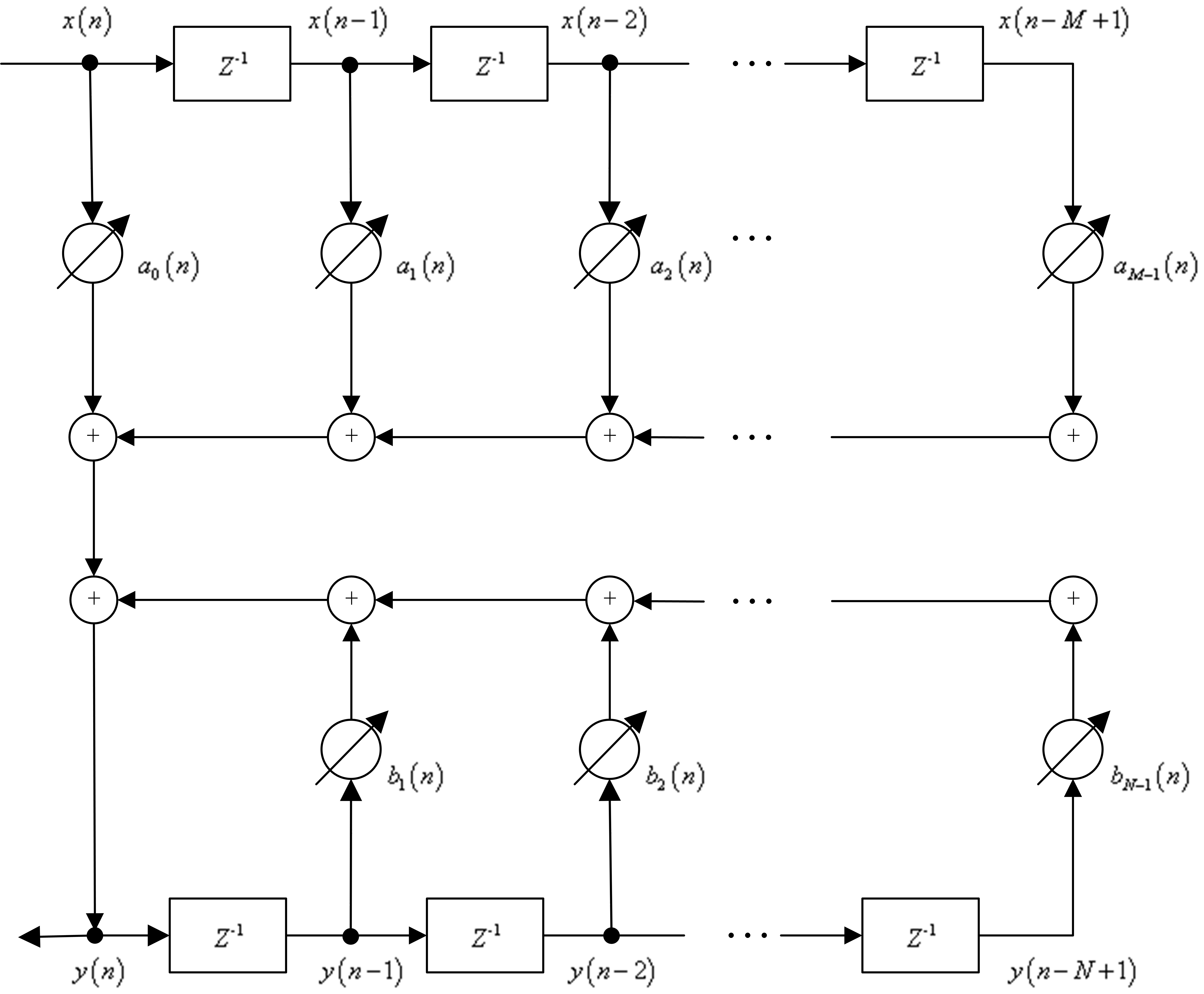}
\caption{Block diagram of an IIR filter.}
\label{fig:IIR} 
\end{figure}

The length of the filter is considered to be $N$, which means that the impulse response of the filter becomes zero after $N$ sampling periods. The input at time $n$ and the $i$th weight coefficient can be represented as 
${x\left( n \right)}$ and ${\omega _i}\left( n \right)$, respectively. The input and weight coefficients of the filter were given by\cite{snyder2000active}:
\begin{equation}
    X(n)=[x(n), x(n-1), \cdots, x(n-N+1)]^T
\end{equation}
\begin{equation}
    W(n)=\left[\omega_0(n), \omega_1(n), \cdots, \omega_{N-1}(n)\right]^T.
\end{equation}
 The output of the filter at time $n$ is given by:
\begin{equation}
    y\left( n \right) = {W^T}\left( n \right)X\left( n \right) = \sum\limits_{i = 0}^{N - 1} {{\omega _i}\left( n \right)x\left( {n - i} \right)}.
    \label{y(n)}
\end{equation}
Based on the expression for the filter's output, the system transfer function can be derived as follows:
\begin{equation}
    H\left( z \right) = \frac{{Y\left( z \right)}}{{X\left( z \right)}} = \sum\limits_{i = 0}^{N - 1} {{\omega _i}\left( n \right){z^{ - i}}}. 
    \label{con:H of FIR}
\end{equation}
IIR filters differ from FIR  filters in that their output signal depends on both the current and previous input and output signals. This recursive nature of IIR filters requires an infinite number of terms to represent their impulse response. The structure of an IIR filter can be demonstrated using the block diagram shown in \autoref{fig:IIR}.

Let $x\left( n \right)$ and $y\left( n \right)$ represent the input and output of the filter at time $n$, respectively. Let ${a_i}\left( n \right)$ and ${b_i}\left( n \right)$ denote the $i$th input coefficient and recursive coefficient of the filter at time $n$. Then, the output of the IIR filter is given by:
\begin{equation}
    y\left( n \right) = \sum\limits_{i = 0}^{M - 1} {{a_i}\left( n \right)x\left( {n - i} \right)}  + \sum\limits_{i = 1}^{N - 1} {{b_i}\left( n \right)y\left( {n - i} \right)}. 
\end{equation}
Therefore, the system transfer function is defined as:
\begin{equation}
    H\left( z \right) = \frac{{Y\left( z \right)}}{{X\left( z \right)}} = \frac{{\sum\limits_{i = 0}^{M - 1} {{a_i}\left( n \right){z^{ - i}}} }}{{1 - \sum\limits_{i = 1}^{N - 1} {{b_i}\left( n \right){z^{ - i}}} }}.
    \label{con:H of IIR}
\end{equation}
By comparing \autoref{con:H of FIR} and \autoref{con:H of IIR}, it can be observed that both FIR and IIR filters have zeros in their transfer functions. However, due to the presence of recursive feedback, IIR filters also have poles. If the poles of an IIR filter lie outside the unit circle, the filter may oscillate and become unstable. Therefore, in practical applications, FIR filters are often preferred for adaptive filtering to ensure the stability of the control system.

\section{Adaptive Filtering Algorithms}\label{Adaptive Filtering Algorithms}

As mentioned above, adaptive filtering algorithms play a crucial role in current feedforward active noise control systems. This section provides a brief overview of the fundamental working principles of the Least Mean Square (LMS) algorithm and its variants used in such applications. Furthermore, essential analysis of the variant algorithms is provided.

\subsection{LMS algorithm}

\begin{figure}[ht]
\centering
\includegraphics{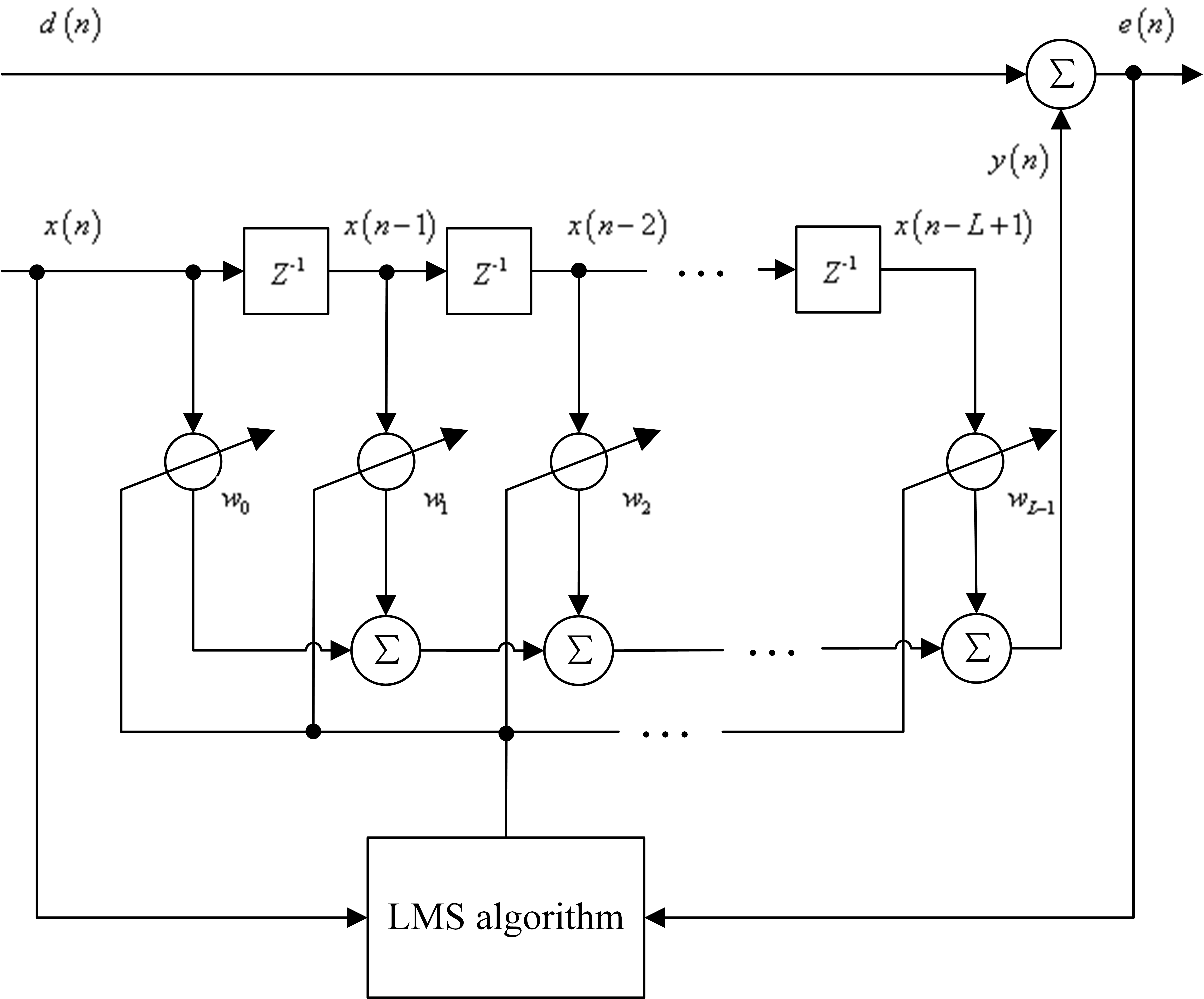}
\caption{Block diagram of an LMS adaptive filter.}
\label{fig:LMS} 
\end{figure}

An adaptive algorithm adjusts the filter coefficients based on specific criteria. The LMS algorithm is the most commonly used adaptive algorithm. It minimizes the MSE, which is the squared difference between the desired signal and the actual output signal\cite{padhi2019cascading}, as the criterion for updating the filter weights. The adaptive filter structure based on the LMS algorithm is illustrated in \autoref{fig:LMS}.

In \autoref{fig:LMS}, $x\left( n \right)$ represents the reference signal, $d\left( n \right)$ represents the desired signal, $e\left( n \right)$ represents the error signal, and $y\left( n \right)$ represents the output signal of the filter. The error signal can be expressed as:
\begin{equation}
    e\left( n \right) = d\left( n \right) - y\left( n \right).
\end{equation}
Combining with \autoref{y(n)}, the error signal is given by:
\begin{equation}
    e\left( n \right) = d\left( n \right) - {W^T}\left( n \right)X\left( n \right).
\end{equation}
According to the minimum mean square error (MMSE) criterion, the cost function of the LMS algorithm is given by:
\begin{equation}
\begin{aligned}
J\left( n \right) & = E\left[ {{e^2}\left( n \right)} \right]\\
& = E\left[ {{{\left( {d\left( n \right) - {W^T}\left( n \right)X\left( n \right)} \right)}^2}} \right]\\
& = E\left[ {\left( {d\left( n \right) - {W^T}\left( n \right)X\left( n \right)} \right){{\left( {d\left( n \right) - {W^T}\left( n \right)X\left( n \right)} \right)}^T}} \right]\\
& = E\left[ {\left( {d\left( n \right) - {W^T}\left( n \right)X\left( n \right)} \right)\left( {d\left( n \right) - {X^T}\left( n \right)W\left( n \right)} \right)} \right]\\
& = E\left[ {{d^2}\left( n \right) + {W^T}\left( n \right)X\left( n \right){X^T}\left( n \right)W\left( n \right) - 2d\left( n \right){W^T}\left( n \right)X\left( n \right)} \right]\\
& = E\left[ {{d^2}\left( n \right)} \right] - 2P{W^T}\left( n \right) + {W^T}\left( n \right) - {W^T}\left( n \right)RW\left( n \right)
\end{aligned}
\end{equation}
where $P$ is the cross-correlation vector between the desired signal and the reference signal, and $R$ is the autocorrelation matrix of the input signal, denoted as $P = E\left[ {\left( {d\left( n \right){X^T}\left( n \right)} \right)} \right]$ and $R = E\left[ {\left( {X\left( n \right){X^T}\left( n \right)} \right)} \right]$, respectively. Therefore, $J\left( n \right)$ is a quadratic function of the weight coefficient vector $W\left( n \right)$ and has a unique minimum point. When the MSE reaches its minimum value, the weight coefficient vector $W\left( n \right)$ achieves the optimal solution ${W^ * }$, which satisfies:
\begin{equation}
    \frac{{\partial J\left( n \right)}}{{\partial W\left( n \right)}}\left| {_{W\left( n \right) = {W^ * }}} \right. = 2RW\left( n \right) - 2P = 0.
\end{equation}
The optimal weight matrix is given by:
\begin{equation}
    W = {R^{ - 1}}P.
\end{equation}
Due to the difficulty of obtaining the autocorrelation matrix $R$ and cross-correlation matrix $P$ in practical applications, iterative approaches are commonly used to calculate the optimal weight matrix $W$. The most widely used iterative method is the steepest descent 
\begin{equation}
    W\left( {n + 1} \right) = W\left( n \right) - \mu \nabla \left( n \right)
    \label{W(n+1)}
\end{equation}
where $\mu$ is the convergence factor, which determines the convergence speed and stability of the system; $\nabla \left( n \right)$ represents the gradient vector at the nth iteration.

Since obtaining an accurate estimate of $\nabla \left( n \right)$ is challenging, it is common to replace the estimation of the mean $E\left[ {{e^2}\left( n \right)} \right]$ with its instantaneous value, resulting in the instantaneous gradient, which is given by:
\begin{equation}
    \hat \nabla  = \frac{{\partial E\left[ {{e^2}\left( n \right)} \right]}}{{\partial W\left( n \right)}} = \frac{{\partial {e^2}\left( n \right)}}{{\partial W\left( n \right)}} = 2e\left( n \right)\frac{{\partial e\left( n \right)}}{{\partial W\left( n \right)}} =  - 2e\left( n \right)X\left( n \right).
\end{equation}
Therefore, \autoref{W(n+1)} can be expressed as:
\begin{equation}
    W\left( {n + 1} \right) = W\left( n \right) + 2\mu e\left( n \right)X\left( n \right).
    \label{LMS W}
\end{equation}
To ensure system stability, the convergence factor $\mu$ is typically maintained within a specific range:
\begin{equation}
    0 < \mu  < \frac{1}{{tr\left[ R \right]}}
\end{equation}
where $tr\left[ R \right]$ is the trace of the autocorrelation matrix $R$, which is equal to the sum of diagonal elements of $R$. It also represents the total power of the input signal. Therefore, the range of $\mu$ can be further expressed as:
\begin{equation}
    0 < \mu  < \frac{1}{{\sum\limits_{i = 0}^{N - 1} {{{\left( {x\left( {n - i} \right)} \right)}^2}} }}
    \label{mu}
\end{equation}
In practical engineering, the total power of the input signal is often known. The range of the convergence factor $\mu$ can be determined based on \autoref{mu}.

\subsection{Single-channel FxLMS Algorithm}
The standard LMS algorithm used in ANC systems can potentially lead to system instability due to the presence of the secondary acoustic path, which results in a time misalignment between the reference signal and the error signal. To address this issue, certain modifications are made to the instantaneous gradient estimation in LMS algorithm.

\begin{figure}[ht]
\centering
\includegraphics{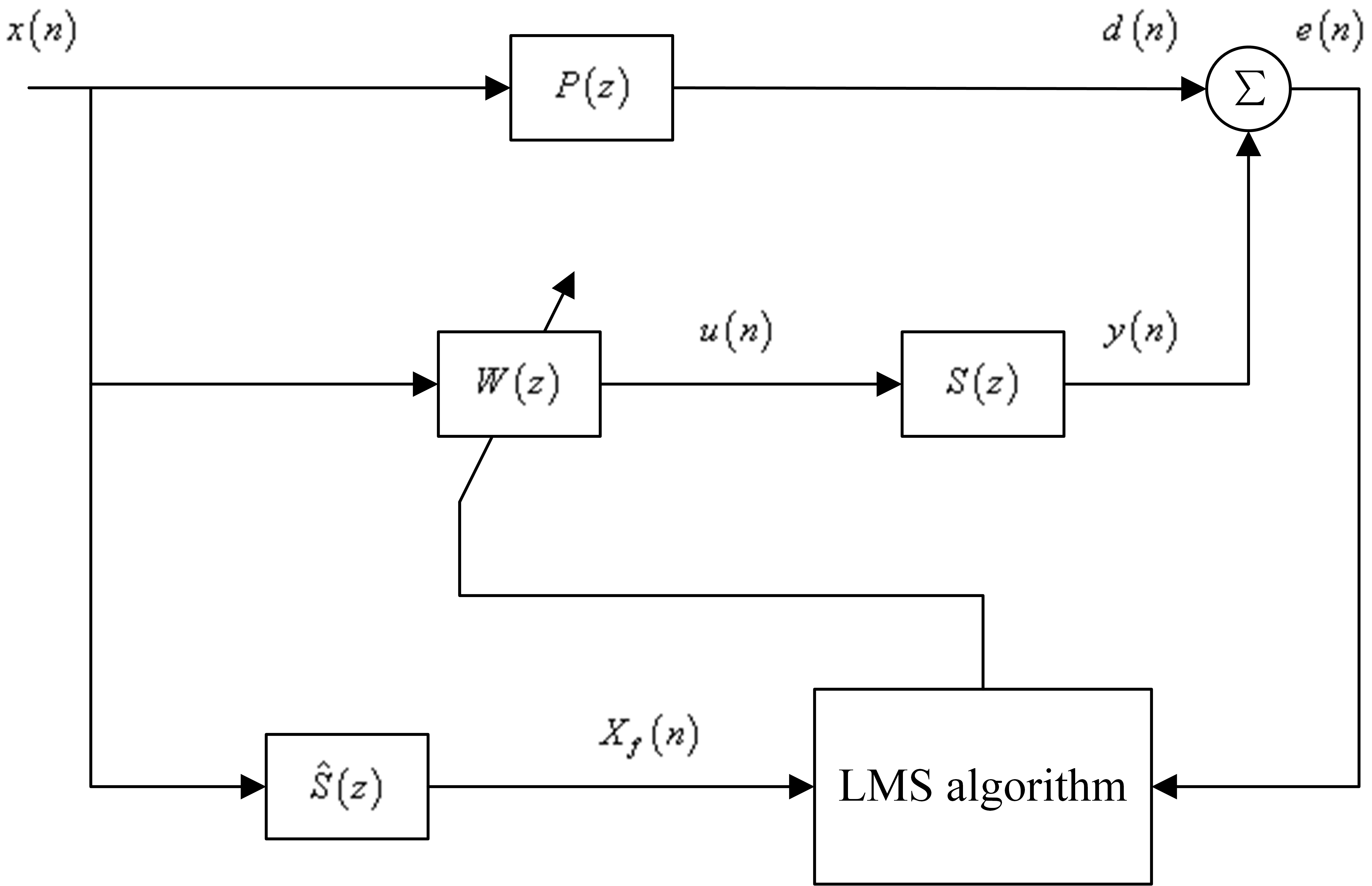}
\caption{Block diagram of an FxLMS algorithm.}
\label{fig:FxLMS} 
\end{figure}

The cost function of the LMS algorithm is given by:
\begin{equation}
    \begin{aligned}
J\left( n \right) & = E\left[ {{e^2}\left( n \right)} \right] \approx {e^2}\left( n \right)\\
 & = {\left( {d\left( n \right) + \sum\limits_{i = 0}^{M - 1} {{h_i}\left( n \right)u\left( {n - i} \right)} } \right)^2}\\
 & = {\left( {d\left( n \right) + \sum\limits_{i = 0}^{M - 1} {{h_i}\left( n \right)\sum\limits_{j = 0}^{N - 1} {{w_j}\left( {n - i} \right)} x\left( {n - i - j} \right)} } \right)^2}
    \end{aligned}
\end{equation}
where $M$ is the length of the secondary path, and ${h_i}\left( n \right)$ represents the $i$th impulse response coefficient of the secondary path at time $n$. Note that the error signal is the sum of the reference signal and the output signal. Therefore, the instantaneous gradient estimate is given by:
\begin{equation}
\begin{aligned}
 \frac{\partial J(n)}{\partial W(n)}& =2 e(n) \frac{\partial y(n)}{\partial W(n)} \\
& =2 e(n) \sum_{i=0}^{M-1} h_i(n) \frac{\partial u(n-i)}{\partial W(n)} \\
& =2 e(n) \sum_{i=0}^{M-1} h_i(n)\left[\begin{array}{c}
\frac{\partial u(n-i)}{\partial w_0(n)} \\
\frac{\partial u(n-i)}{\partial w_1(n)} \\
\vdots \\
\frac{\partial u(n-i)}{\partial w_{N-1}(n)}
\label{J(nn)}
\end{array}\right] \\
\end{aligned}
\end{equation}
Assuming that the filter weight coefficients are updated very slowly over a certain period of time, meaning that the step size factor is very small, it can be considered that:
\begin{equation}
    \frac{{\partial u\left( {n - i} \right)}}{{\partial {w_j}\left( n \right)}} \approx \frac{{\partial u\left( {n - i} \right)}}{{\partial {w_j}\left( {n - i} \right)}},j = 0,1, \cdot  \cdot  \cdot ,N - 1
\end{equation}
Therefore, \autoref{J(nn)} can be expressed as:
\begin{equation}
    \begin{aligned}
        \begin{array}{l}
\frac{{\partial J(n)}}{{\partial W(n)}} \approx  2e(n)\sum\limits_{i = 0}^{M - 1} {{h_i}} (n)\left[ {\begin{array}{*{20}{c}}
{\frac{{\partial u(n - i)}}{{\partial {w_0}(n - i)}}}\\
{\frac{{\partial u(n - i)}}{{\partial {w_1}(n - i)}}}\\
 \vdots \\
{\frac{{\partial u(n - i)}}{{\partial {w_{N - 1}}(n - i)}}}
\end{array}} \right]\\
  = 2e(n)\left[ {\begin{array}{*{20}{c}}
{\sum\limits_{i = 0}^{M - 1} {{h_i}} (n)\frac{{\partial u(n - i)}}{{\partial {w_0}(n - i)}}}\\
{\sum\limits_{i = 0}^{M - 1} {{h_i}} (n)\frac{{\partial u(n - i)}}{{\partial {w_1}(n - i)}}}\\
 \vdots \\
{\sum\limits_{i = 0}^{M - 1} {{h_i}} (n)\frac{{\partial u(n - i)}}{{\partial {w_{N - 1}}(n - i)}}}\\
\end{array}} \right]\\
 = 2e\left( n \right){X_f}\left( n \right)
\end{array}
    \end{aligned}
\end{equation}
where ${X_f}\left( n \right)$ is defined as:
\begin{equation}
    \begin{aligned}
        \begin{array}{l}
{X_f}\left( n \right) = {\left[ {{x_f}\left( n \right),{x_f}\left( {n - 1} \right), \cdot  \cdot  \cdot ,{x_f}\left( {n - N + 1} \right)} \right]^T}\\
{x_f}\left( n \right) = \sum\limits_{i = 0}^{M - 1} {{h_i}\left( n \right)x\left( {n - i} \right)} .
\end{array}
    \end{aligned}
\end{equation}
 The update equation for the weight coefficients in the FxLMS algorithm is given by\cite{burgess1981active}:
 \begin{equation}
     W\left( {n + 1} \right) = W\left( n \right) - 2\mu e\left( n \right){X_f}\left( n \right)
     \label{FxLMS W}
 \end{equation}
 By comparing \autoref{LMS W} and \autoref{FxLMS W}, it can be observed that ${X_f}\left( n \right)$ replaces the reference input signal vector $X\left( n \right)$ used in the LMS algorithm. It represents the reference input signal vector $X\left( n \right)$ filtered through the secondary path. The FxLMS algorithm is depicted in the following \autoref{fig:FxLMS}.

The execution steps of the FxLMS algorithm can be summarized as follows:
\begin{enumerate}
    \item The reference microphone collects the reference signal $x\left( n \right)$, while the error microphone collects the error signal $e\left( n \right)$.
    \item Compute the filtered output signal $u\left( n \right)$:
    \begin{equation}
        u\left( n \right) = {W^T}\left( n \right)X\left( n \right) = \sum\limits_{i = 0}^{N - 1} {{w_i}\left( n \right)x\left( {n - i} \right)} 
    \end{equation}
    \item The output signal $u\left( n \right)$ drives the secondary loudspeaker.
    \item Compute the filtered-X signal ${x_f}\left( n \right)$:
    \begin{equation}
        {x_f}\left( n \right) = \sum\limits_{i = 0}^{M - 1} {{s_i}\left( n \right)x\left( {n - i} \right)} 
    \end{equation}
    \item Update the weight coefficients of the adaptive filter $W\left( z \right)$ using the FxLMS algorithm:
    \begin{equation}
        W\left( {n + 1} \right) = W\left( n \right) - 2\mu e\left( n \right){X_f}\left( n \right)
    \end{equation}
    \item Repeat the above steps until the error signal $e\left( n \right)$ meets the desired criteria.
\end{enumerate}

\subsection{Convergence Analysis on Single-channel FxLMS}
Assuming ${W_{opt}}$ is the optimal weight coefficient vector and ${W_{opt}}{X_f} \approx  - d\left( n \right)$, the deviation vector $\varepsilon \left( n \right)$ is defined as\cite{haykin2002adaptive}:
\begin{equation}
    \begin{array}{l}
\varepsilon \left( n \right) = {W_{opt}} - W\left( n \right)\\
\varepsilon \left( {n + 1} \right) = \varepsilon \left( n \right) + \mu \psi \left( n \right){X_f}\left( n \right)\\
\psi \left( n \right) = q\left( {x\left( n \right)} \right)e\left( n \right).
\end{array}
\end{equation}
 The mean square deviation value is defined as:
 \begin{equation}
     \begin{array}{l}
\delta \left( n \right) = E\left\{ {{{\left\| {\varepsilon \left( n \right)} \right\|}^2}} \right\}\\
E\left\{ {{{\left\| {\varepsilon \left( {n + 1} \right)} \right\|}^2}} \right\} = E\left\{ {{{\left\| {\varepsilon \left( n \right) + \mu \psi \left( n \right){X_f}\left( n \right)} \right\|}^2}} \right\}.
\end{array}
 \end{equation}
Due to:
\begin{equation}
\begin{aligned}   
{\left\| {\varepsilon \left( n \right) + \mu \psi \left( n \right){X_f}\left( n \right)} \right\|^2} & = \left\langle {\varepsilon \left( n \right) + \mu \psi \left( n \right){X_f}\left( n \right),\varepsilon \left( n \right) + \mu \psi \left( n \right){X_f}\left( n \right)} \right\rangle \\
 & = {\left\| {\varepsilon \left( n \right)} \right\|^2} + {\left\| {\mu \psi \left( n \right){X_f}\left( n \right)} \right\|^2} \\ & + 2\mu \psi \left( n \right){\varepsilon ^T}\left( n \right){X_f}\left( n \right)\\
 {\varepsilon ^T}\left( n \right){X_f}\left( n \right) & = {\left( {{W_{opt}} - W\left( n \right)} \right)^T}{X_f}\left( n \right) \approx  - d\left( n \right) - y\left( n \right) =  - e\left( n \right),
\end{aligned}
\end{equation}
the difference of MSE is given by:
\begin{equation}
    \begin{aligned}
\Delta \left( n \right) & = \delta \left( {n + 1} \right) - \delta \left( n \right)\\
 & = {\mu ^2}E\left\{ {{\psi ^2}\left( n \right){{\left\| {{X_f}\left( n \right)} \right\|}^2}} \right\} + 2\mu E\left\{ {\psi \left( n \right){\varepsilon ^T}\left( n \right){X_f}\left( n \right)} \right\}\\
 & = {\mu ^2}E\left\{ {{\psi ^2}\left( n \right){{\left\| {{X_f}\left( n \right)} \right\|}^2}} \right\} - 2\mu E\left\{ {\psi \left( n \right)e\left( n \right)} \right\}.
    \end{aligned}
\end{equation}
To ensure algorithm convergence, the condition for the step size factor $\mu$ is given by:
\begin{equation}
    0 < \mu  < \frac{{2E\left\{ {\psi \left( n \right)e\left( n \right)} \right\}}}{{E\left\{ {{\psi ^2}\left( n \right){{\left\| {{X_f}\left( n \right)} \right\|}^2}} \right\}}}
\end{equation}

\subsection{Multichannel FxLMS Algorithm}
From the perspective of controller structure, the basic principles of constructing multichannel adaptive algorithms are the same.The significant difference lies in the significantly larger computational complexity of multichannel adaptive algorithms compared to single-channel algorithms. Moreover, in multichannel systems, there exists mutual coupling between different channels, making the stability of the algorithm more challenging to ensure.

\begin{figure}[ht]
\centering
\includegraphics{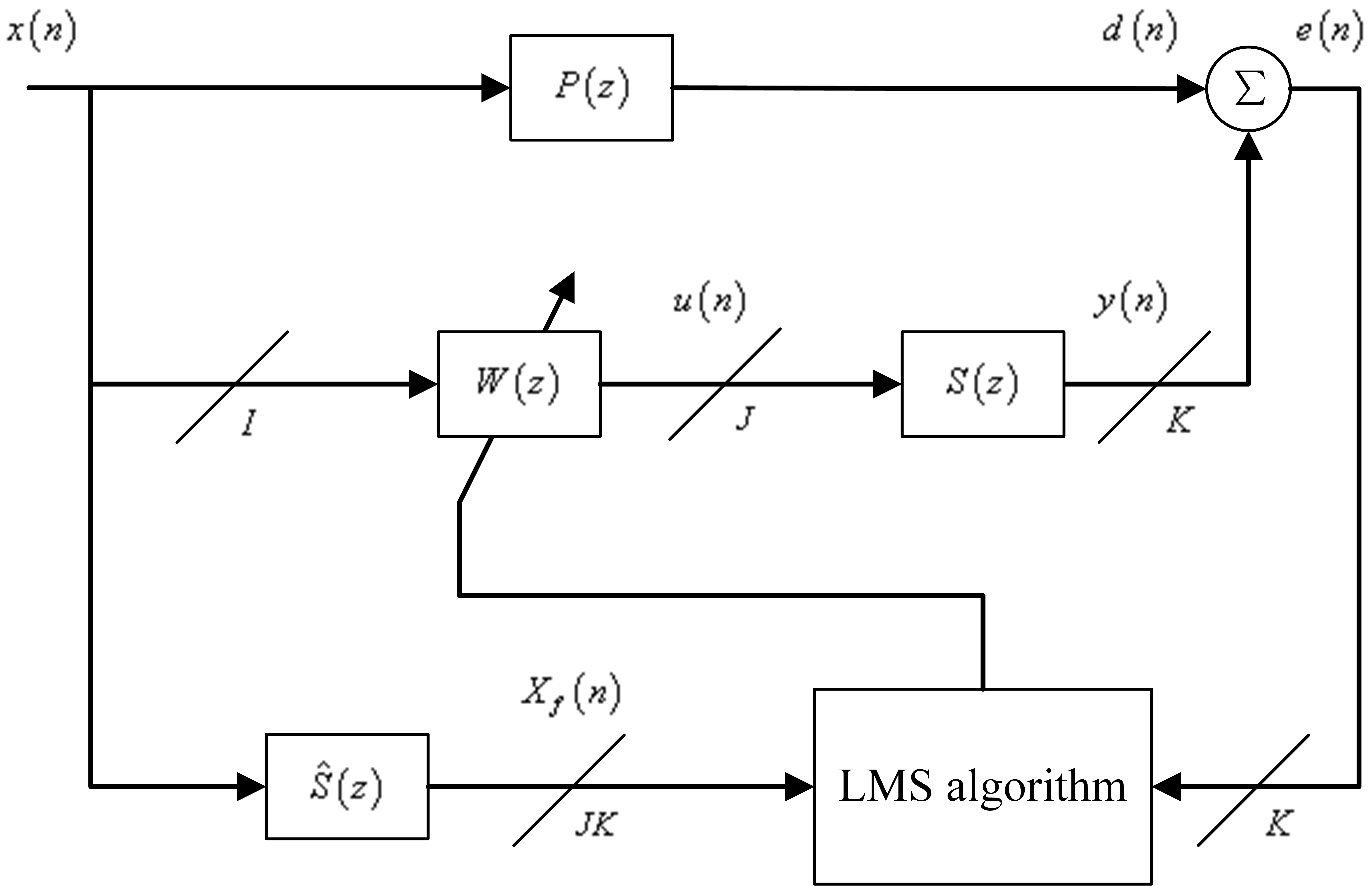}
\caption{Block diagram of the multichannel FxLMS algorithm}
\label{fig:MCFxLMS} 
\end{figure}

\autoref{fig:MCFxLMS} shows the block diagram of the multichannel FxLMS algorithm, which has $I$ reference microphones, $J$ secondary sources, and $K$ error microphones.

The controller requires $I \times J$ adaptive FIR filters, assuming each filter has a length of $L$. The weight coefficient matrix of this controller is given by\cite{kuo1996active}:
\begin{equation}
\mathbf{W}(n)=\left[\begin{array}{cccc}
\mathbf{w}^{(1,1)}(n) & \mathbf{w}^{(1,2)}(n) & \cdots & \mathbf{w}^{(1, J)}(n) \\
\mathbf{w}^{(2,1)}(n) & \mathbf{w}^{(2,2)}(n) & \cdots & \mathbf{w}^{(2, J)}(n) \\
\vdots & \vdots & \ddots & \vdots \\
\mathbf{w}^{(I, 1)}(n) & \mathbf{w}^{(1,2)}(n) & \cdots & \mathbf{w}^{(I, J)}(n)
\end{array}\right]
\end{equation}
where the weight coefficient vector of the $\left( {i,j} \right)$th filter is given by:
\begin{equation}
    \begin{aligned}
{w^{\left( {i,j} \right)}}\left( n \right) & = {\left[ {w_0^{\left( {i,j} \right)}\left( n \right),w_1^{\left( {i,j} \right)}\left( n \right), \cdot  \cdot  \cdot ,w_{L - 1}^{\left( {i,j} \right)}\left( n \right)} \right]^T}\\
i & = 1,2, \cdot  \cdot  \cdot ,I\\
j & = 1,2, \cdot  \cdot  \cdot ,J.
 \end{aligned}
\end{equation}
The matrix of the reference signal vector is given by;
\begin{equation}
    X\left( n \right) = {\left[ {{x^{\left( 1 \right)}}\left( n \right),{x^{\left( 2 \right)}}\left( n \right), \cdot  \cdot  \cdot ,{x^{\left( I \right)}}\left( n \right)} \right]^T}
\end{equation}
where the $i$th reference signal vector is given by:
\begin{equation}
    \begin{aligned}
{x^{\left( i \right)}}\left( n \right) &= {\left[ {{x_i}\left( n \right),{x_i}\left( {n - 1} \right), \cdot  \cdot  \cdot ,{x_i}\left( {n - L + 1} \right)} \right]^T}\\
i &= 1,2, \cdot  \cdot  \cdot ,I.
\label{xi}
\end{aligned}
\end{equation}
In \autoref{xi}, ${x_i}\left( n \right)$ represents the output signal of the $i$th reference sensor at time $n$. The secondary sound source signal output by the controller is given by:
\begin{equation}
    y\left( n \right) = {\left[ {{y^{\left( 1 \right)}}\left( n \right),{y^{\left( 2 \right)}}\left( n \right), \cdot  \cdot  \cdot ,{y^{\left( J \right)}}\left( n \right)} \right]^T}
\end{equation}
where the output signal of the $j$th secondary sound source at time $n$ is given by:
\begin{equation}
    \begin{aligned}
{y^{\left( j \right)}}\left( n \right) &= \sum\limits_{i = 1}^I {{{\left( {{w^{\left( {i,j} \right)}}\left( n \right)} \right)}^T}{x^{\left( i \right)}}\left( n \right)} \\
 &= \sum\limits_{i = 1}^I {\sum\limits_{l = 0}^{L - 1} {w_l^{\left( {i,j} \right)}\left( n \right){x_i}\left( {n - l} \right)} } \\
j & = 1,2, \cdot  \cdot  \cdot ,J.
    \end{aligned}
\end{equation}
The error signal vector is given by;
\begin{equation}
    e\left( n \right) = {\left[ {{e^{\left( 1 \right)}}\left( n \right),{e^{\left( 2 \right)}}\left( n \right), \cdot  \cdot  \cdot ,{e^{\left( K \right)}}\left( n \right)} \right]^T}
\end{equation}
where ${e^{\left( K \right)}}\left( n \right)$ is the error signal collected at the $k$th error microphone. The cost function of the multichannel FxLMS algorithm is given by;
\begin{equation}
    J\left( n \right) = {\sum\limits_{k = 1}^K {\left( {{e^{\left( k \right)}}\left( n \right)} \right)} ^2} = {\left( {{e^{\left( k \right)}}\left( n \right)} \right)^T}{e^{\left( k \right)}}\left( n \right).
\end{equation}
Similar to the derivation process in the single-channel FxLMS algorithm, the recursive update formula for the weight coefficients of the $\left( {i,j} \right)$th filter in the controller can be derived as follows:
\begin{equation}
    \begin{aligned} 
{w^{\left( {i,j} \right)}}\left( {n + 1} \right) &= {w^{\left( {i,j} \right)}}\left( n \right) - \mu \sum\limits_{k = 1}^K {{e^{\left( k \right)}}\left( n \right){f_x}^{\left( {i,j,k} \right)}\left( n \right)} \\
i &= 1,2, \cdot  \cdot  \cdot ,I\\
j &= 1,2, \cdot  \cdot  \cdot ,J
    \end{aligned}
\end{equation}
where the $\left( {i,j,k} \right)$th fitler-x signal vector is given by:
\begin{equation}
    {f_x}^{\left( {i,j,k} \right)}\left( n \right) = {\left[ {{f_{{x_{i,j,k}}}}\left( n \right),{f_{{x_{i,j,k}}}}\left( {n - 1} \right), \cdot  \cdot  \cdot ,{f_{{x_{i,j,k}}}}\left( {n - L + 1} \right)} \right]^T}
    \label{fx}
\end{equation}
In \autoref{fx}, ${f_{{x_{i,j,k}}}}\left( n \right)$ is the $\left( {i,j,k} \right)$th fitler-x signal at time $n$ given by:
\begin{equation}
    {f_{{x_{i,j,k}}}}\left( n \right) = \sum\limits_{m = 1}^{M - 1} {\hat h{s_m}^{\left( {j,k} \right)}} \left( n \right){x_i}\left( {n - m} \right)
\end{equation}
where ${\hat h{s_m}^{\left( {j,k} \right)}} \left( n \right)$ represents the estimate of the impulse response of the $\left( {j,k} \right)$th secondary path. Assuming that all $J \times K$ secondary paths are designed using FIR filters of length M, the impulse response estimate can be expressed as:
\begin{equation}
    \begin{aligned}
\hat h{s^{\left( {j,k} \right)}}\left( n \right) &= {\left[ {\hat h{s_0}^{\left( {j,k} \right)}\left( n \right),\hat h{s_1}^{\left( {j,k} \right)}\left( n \right), \cdot  \cdot  \cdot ,\hat h{s_{M - 1}}^{\left( {j,k} \right)}\left( n \right)} \right]^T}\\
j &= 1,2, \cdot  \cdot  \cdot ,J\\
k &= 1,2, \cdot  \cdot  \cdot ,K.
    \end{aligned}
\end{equation}

\subsection{Computational Complexity of Multichannel FxLMS Algorithm}
The computation time of each iteration loop in an adaptive control algorithm is equal to the number of instructions required multiplied by the instruction cycle of the DSP processor. The most typical operation in adaptive control algorithms is the multiplication and accumulation (MAC) operation\cite{reddy2008fast}. Therefore, the number of MAC operations performed in one loop is often used as a metric to measure the computational complexity of the algorithm. By carefully studying the implementation process of the algorithm, it is possible to estimate the average number of MAC operations that need to be completed within each sampling period.

In addition, the amount of memory required to perform the algorithm's operations is an important metric for evaluating algorithm complexity. DSP computation involves a significant amount of data access. In general, access to data within the DSP chip can be achieved within one instruction cycle. However, accessing data from external random access memory (RAM) takes more time.

If an algorithm requires a large amount of memory, implementing that algorithm will significantly increase the computation time and also increase the cost of the controller.When designing and selecting algorithms, it is important to consider their data storage requirements. Wherever possible, it is preferable to choose algorithms with lower memory requirements while still meeting the application's requirements. Generally, the storage requirements of an algorithm are proportional to its computational complexity.

In a McANC system, there are $I$ reference sensors, $J$ secondary sound sources, and $K$ error sensors. The controller is a length-L adaptive transversal FIR filter, and the secondary path is modeled as an FIR filter with length M. For the multichannel FxLMS algorithm, the computational steps and complexity required in one sampling period are summarized in \autoref{complexity}.

\begin{table}[ht]
\centering
\caption{The computational steps and complexity for the multichannel FxLMS algorithm.}
\label{complexity}
\begin{tabular}[t]{lcc}
\toprule
Steps&Formulas&Complexity\\
\midrule
1&$\begin{array}{r}
y^{(j)}(n)=\sum\limits_{i=1}^l \sum\limits_{l=0}^{L-1} w_l^{(i, j)}(n) x^{(i)}(n-l) \\
j=1,2, \cdots, J
\end{array}$&$IJL$\\
2&
$\begin{array}{r} f x^{(i, j, k)}(n)  =\sum\limits_{m=1}^{M-1} \hat{h} s_m^{(j, k)}(n) x^{(i)}(n-m) \\ j =1,2, \cdots, J \\ k=1,2, \cdots, K \end{array}$
&$IJKM$\\
3&$\begin{array}{r}w_l^{(i, j)}(n+1)=w_l^{(i, j)}(n)-\mu \sum\limits_{k=1}^K e^{(k)}(n) f x^{(i, j, k)}(n-l) \\ i=1,2, \cdots, I \\ j=1,2, \cdots, J  \\ l=0,1, \cdots, L-1\end{array}$&$IJKL+K$\\
\bottomrule
\end{tabular}
\end{table}%

The average number of MAC operations required per iteration in the algorithm is given by:
\begin{equation}
    MA{C_{FxLMS}} = \left( {IJK + IJ} \right)L + IJKM + K.
\end{equation}

If a standard McANC system is be considered with $N = I = J = K,L = M$, the computational complexity is given by:
\begin{equation}
    MAC_{FxLMS}^s = 2L{N^3} + L{N^2} + N
\end{equation}

As the number of channels in the control system increases, the computational complexity of the multichannel FxLMS algorithm will increase rapidly with the cube of the number of channels.

\section{Summary}
This chapter presented a comprehensive approach to ANC by explaining various aspects of the acoustic mechanism, adaptive control, adaptive filter, and adaptive filtering algorithms.

Section 3.1 delved into the fundamental understanding of how ANC worked by analyzing the acoustic mechanism involved. It discussed the principles of sound propagation, interference, and cancellation, highlighting the importance of phase and amplitude relationships in achieving effective noise reduction.

Section 3.2 introduced the main parts of adaptive control in ANC systems. It discussed the importance of adaptive algorithms as the focus of this research.  

Section 3.3 provided an in-depth exploration of adaptive filters. Within this section, both FIR and IIR filters were discussed in detail. The characteristics, advantages, and challenges associated with each type of filter were explored.

Section 3.4 delved into specific adaptive filtering algorithms used in ANC, including the LMS algorithm, single-channel FxLMS algorithm, and multichannel FxLMS algorithm. It explored the theoretical foundations, highlighting their strengths, limitations, and areas for improvement.
\end{spacing}
\newpage


\chapter{Simulation Results}
\begin{spacing}{1.5}
\setlength{\parskip}{0.3in}
This chapter focuses on evaluating the noise reduction performance of the FxLMS algorithm and the pre-trained control filter using real-world noise signals as the reference input. The effectiveness of both algorithms is examined through extensive noise reduction simulations and comparisons.

Through the simulations, we aim to investigate the ability of each method to attenuate the real-world noise and assess their respective strengths and weaknesses. The results obtained from the simulations will provide valuable insights into the comparative performance of the FxLMS algorithm and the pre-trained control filter in reducing real-world noise.

\section{Noise Reduction Simulation}

In this section, we present the practical implementation of the FxLMS algorithm and the pre-trained control filter for reducing real-world noise. To conduct our experiments, we generated two distinct noise signals by combining and mixing an aircraft noise signal with a frequency range of 50 Hz to 14000 Hz and a traffic noise signal with a frequency range of 40 Hz to 1400 Hz. These synthesized noise signals serve as the reference inputs for evaluating the performance of the FxLMS algorithm and the pre-trained control filter in noise attenuation.

\subsection{Combined Noise Cancellation}

\begin{figure}[H]
\centering
\includegraphics[width=11.5cm]{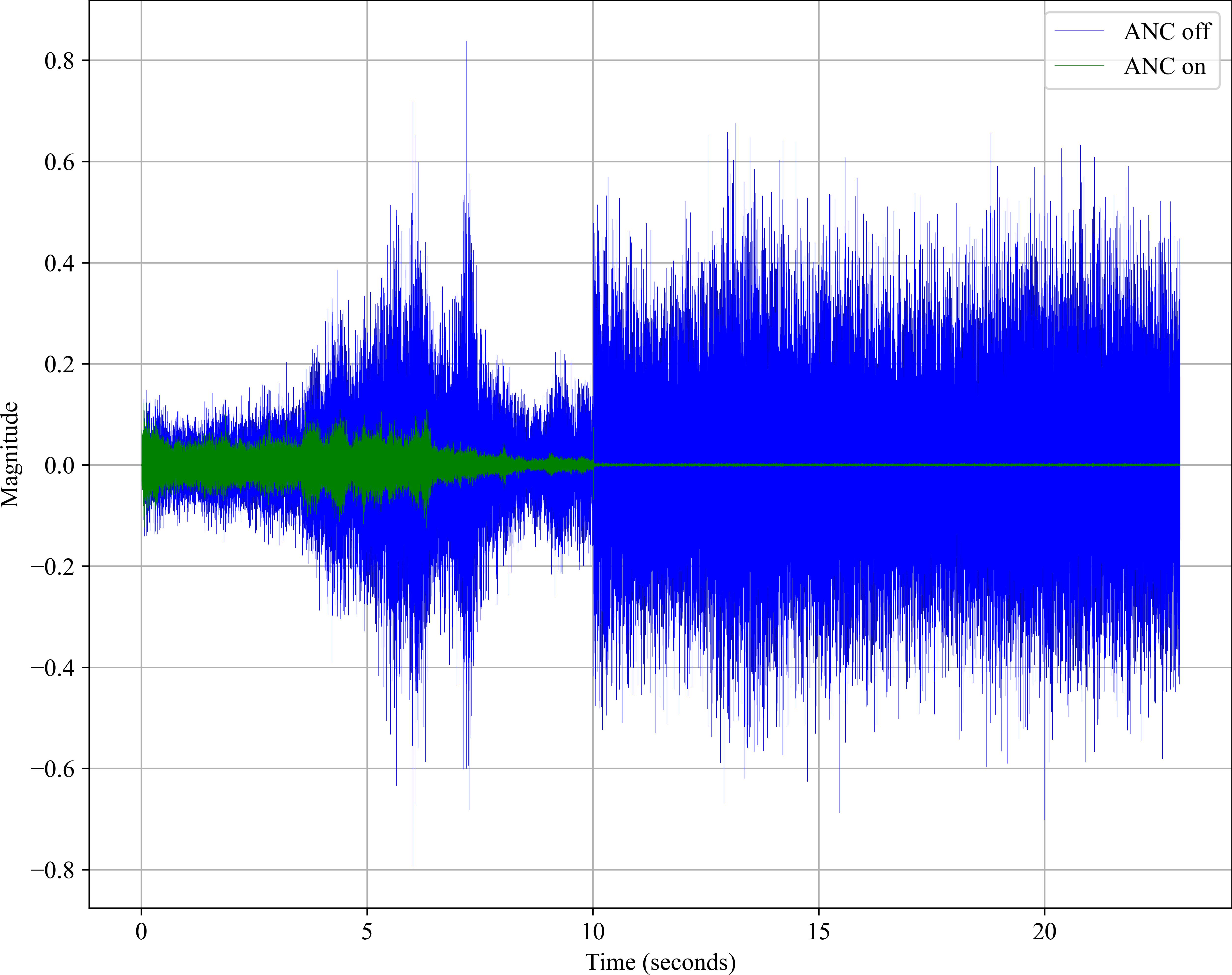}
\caption{Error signals and reference signals of FxLMS algorithm on combined noise.}
\label{fig:connect_Fxlms} 
\end{figure}

\begin{figure}[H]
\centering
\includegraphics[width=11.5cm]{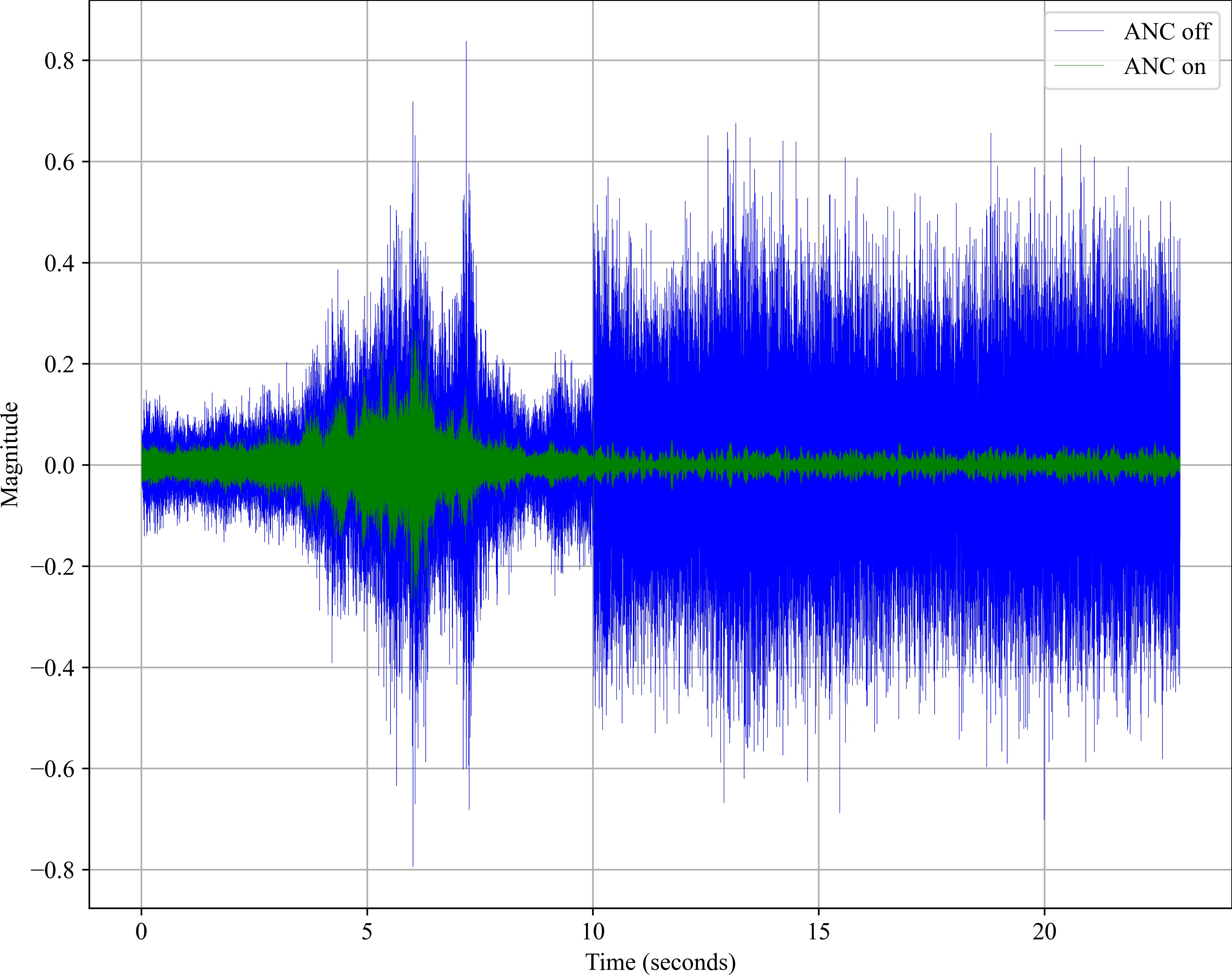}
\caption{Error signals and reference signals of pre-trained control filter on the combined noise.}
\label{fig:Connect_Fix} 
\end{figure}

\begin{figure}[H]
\centering
\includegraphics[width=11.5cm]{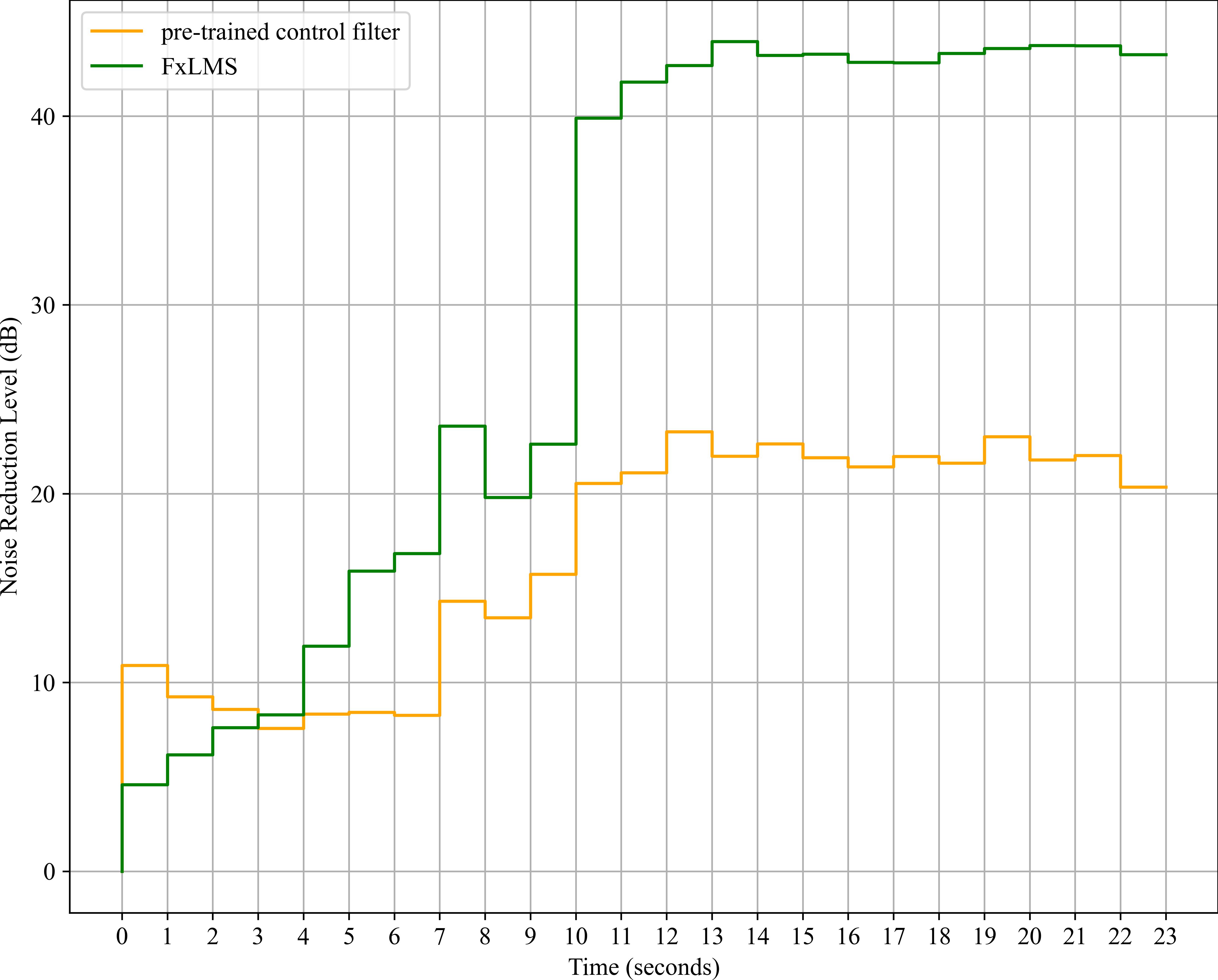}
\caption{Average noise reduction level per one-second interval for the combined noise.}
\label{fig:Noise_Reduction_Level_connected} 
\end{figure}

To simulate noise reduction, we generated a synthesized noise signal by combining aircraft noise and traffic noise. This synthesized noise signal was used as the reference signal for the McANC system. The time-domain response result of the combined noise using the FxLMS algorithm is shown in \autoref{fig:connect_Fxlms}. The time-domain response result of the combined noise using the pre-trained control filter is shown in \autoref{fig:Connect_Fix}. \autoref{fig:Noise_Reduction_Level_connected} illustrates the noise reduction levels on the combined noise achieved by both methods.

From the results, it can be observed that the FxLMS algorithm can effectively track and respond to noise. During the first 3 seconds of the McANC process, the pre-trained control filter outperforms the FxLMS algorithm in terms of noise reduction performance. In particular, during the first second, the average noise reduction levels of FxLMS and pre-trained control filter are 4 dB and 11 dB, respectively. Over the next two seconds, the difference between the two methods gradually decreases. 

However, after 3 seconds, the FxLMS algorithm achieves a significantly better average noise reduction level compared to the pre-trained control filter. At the 23rd second, the difference between the average noise reduction levels of FxLMS and the pre-trained control filter reaches its peak at 23 dB, with FxLMS exhibiting a significantly higher reduction compared to the pre-trained control filter. The FxLMS algorithm exhibits a higher signal-to-noise ratio (SNR) of 21 dB. In contrast, the pre-trained control filter achieves an SNR of 15 dB.

It is worth noting that during the period from 10 to 11 seconds, which represents a transition from one type of noise to another, the FxLMS algorithm exhibits a faster response to the noise compared to the pre-trained control filter. 

\begin{figure}[H]
\centering
\includegraphics[width=11.5cm]{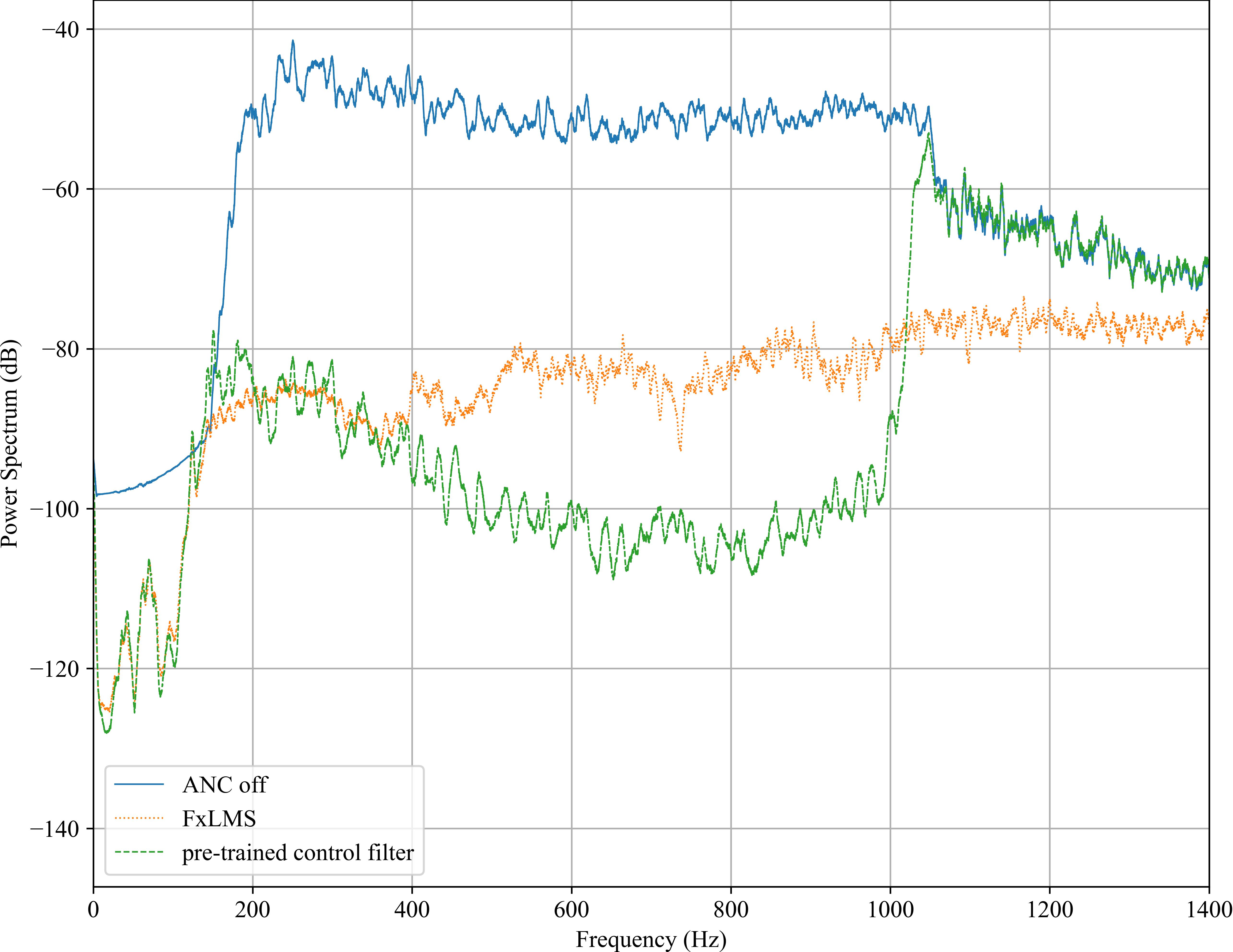}
\caption{Power spectrum of the error signal for FxLMS and pre-trained control filter in the presence of the combined noise.}
\label{fig:Power_Spectrum_combined} 
\end{figure}

\begin{figure}[H]
\centering
\includegraphics[width=11.5cm]{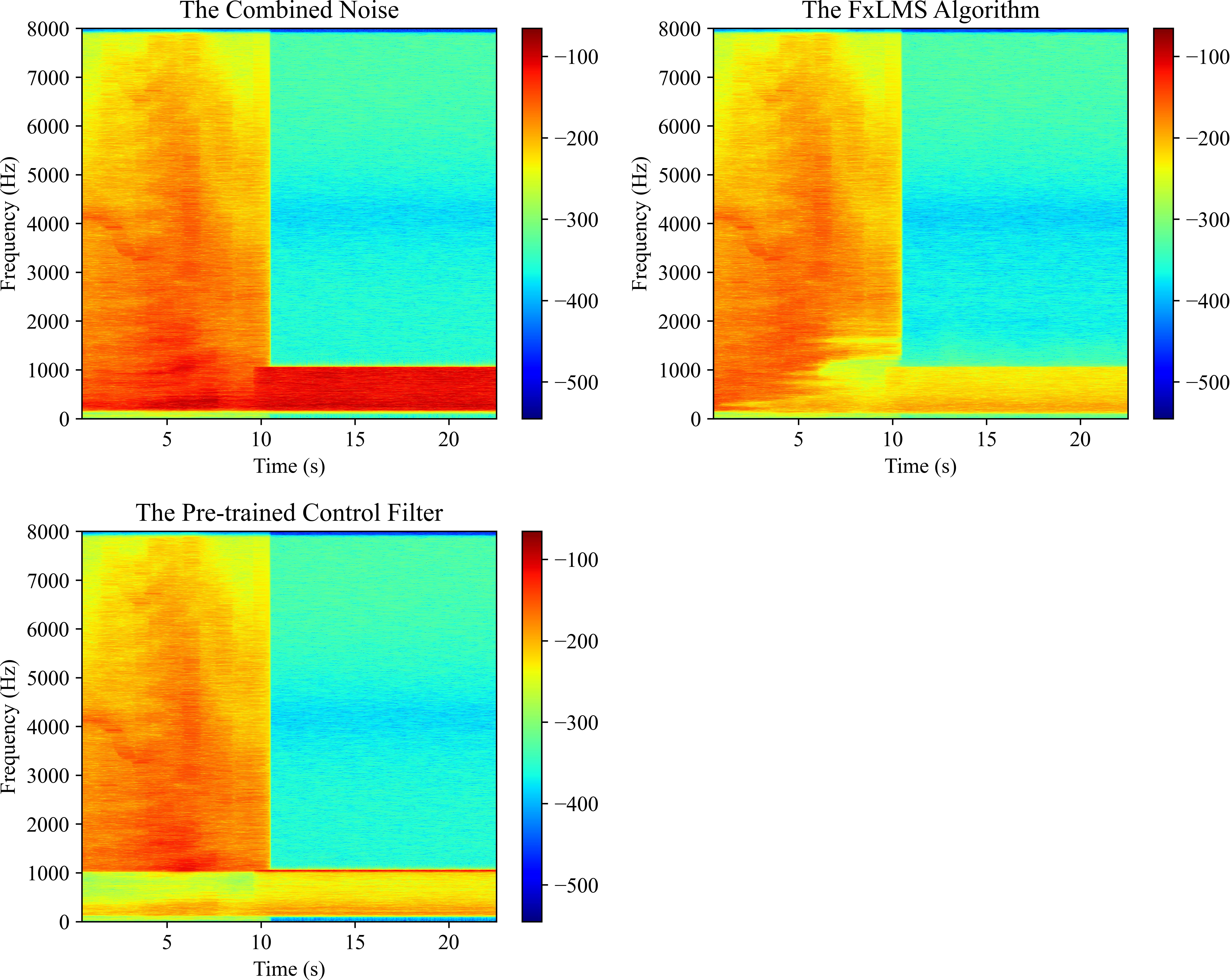}
\caption{Spectrogram of the error signal for FxLMS and pre-trained control filter in the presence of the combined noise.}
\label{fig:Spectrogram_combined} 
\end{figure}

For the combined noise, Both the FxLMS algorithm and the pre-trained control filter exhibit significant attenuation of the noise across different frequency bands, as illustrated in \autoref{fig:Power_Spectrum_combined} and \autoref{fig:Spectrogram_combined}. The power spectrum of the FxLMS algorithm shows a smoother and more consistent reduction in noise across the entire frequency spectrum. The pre-trained control filter's power spectrum exhibits more fluctuations and variations in noise reduction levels at different frequencies.

The power spectrum of the pre-trained control filter exhibits better performance in the low-frequency range, specifically between 400 Hz and 1000 Hz. In this frequency range, the pre-trained control filter demonstrates superior noise reduction capabilities compared to the FxLMS algorithm. On the other hand, the FxLMS algorithm excels in the high frequency range, especially above 1000 Hz, showing better noise suppression abilities than the pre-trained control filter.

From the spectrogram, it can be observed that starting from the third second, the FxLMS algorithm converges and exhibits effective noise suppression across the entire frequency range. In contrast, the pre-trained control filter only has a noticeable impact on low frequency noise and shows minimal influence on high-frequency noise throughout the duration.

\subsection{Mixed Noise Cancellation}

\begin{figure}[H]
\centering
\includegraphics[width=11.5cm]{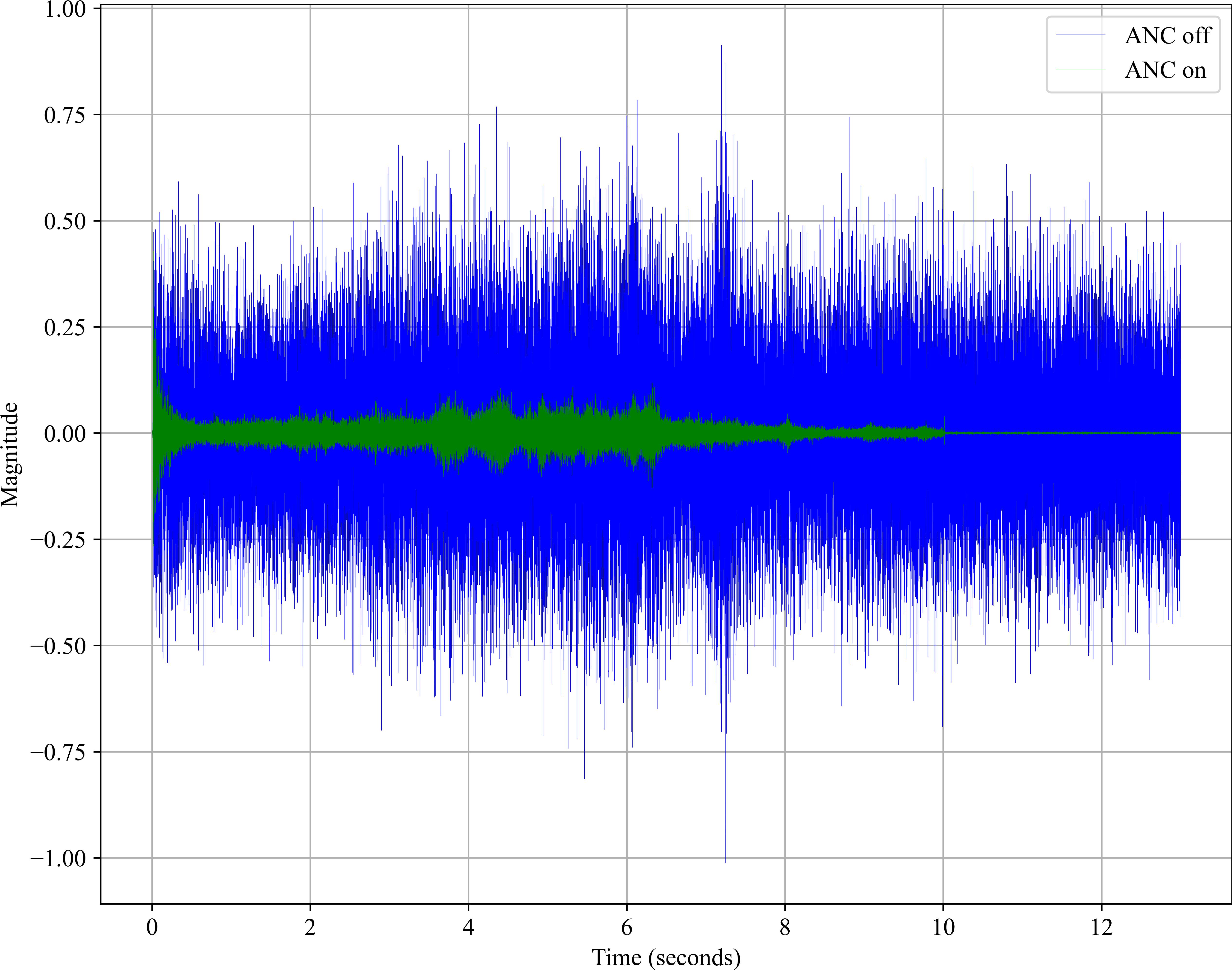}
\caption{Error signals and reference signals of FxLMS algorithm on the mixed noise.}
\label{fig:Mix-FxLMS} 
\end{figure}

\begin{figure}[H]
\centering
\includegraphics[width=11.5cm]{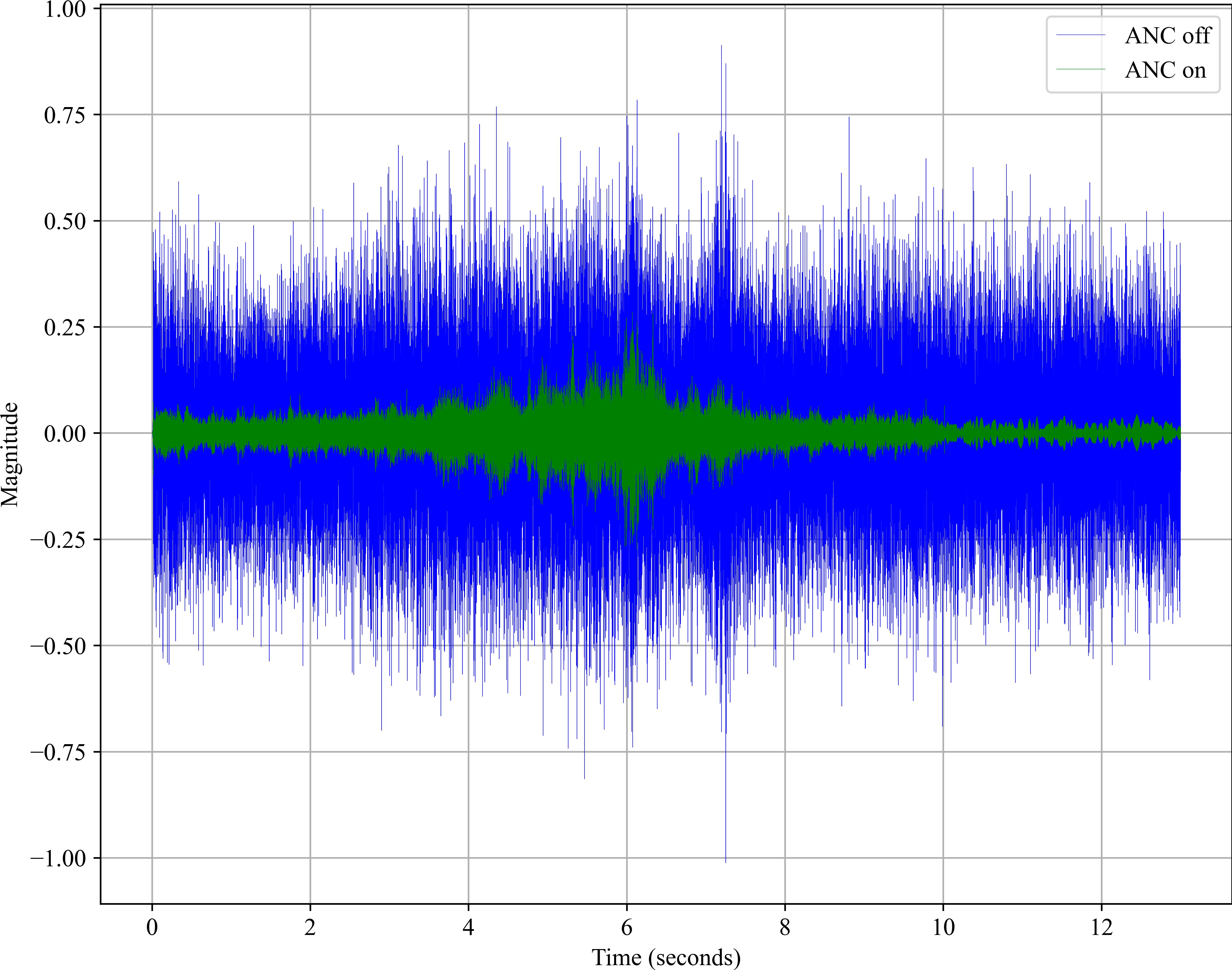}
\caption{Error signals and reference signals of pre-trained control filter on mixed noise.}
\label{fig:Mix_Fix} 
\end{figure}

\begin{figure}[H]
\centering
\includegraphics[width=11.5cm]{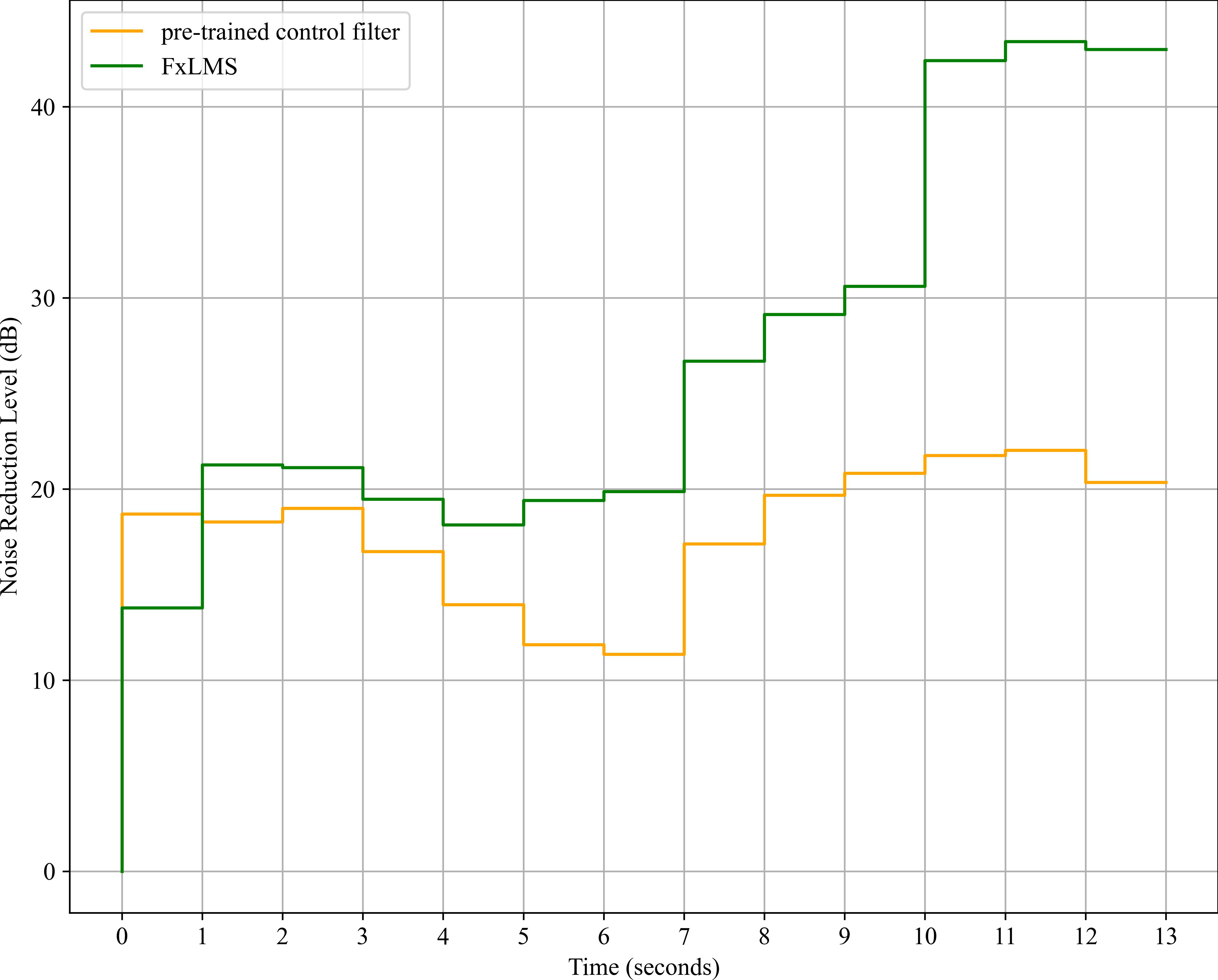}
\caption{Average noise reduction level per one-second interval for the mixed noise.}
\label{fig:Noise_Reduction_Level_mixted} 
\end{figure}

In addition to the combining case, we conducted experiments with a synthesized noise by mixing the aircraft noise and traffic noise. The performance of the FxLMS algorithm on the mixed noise is presented in \autoref{fig:Mix-FxLMS}, while the performance of the pre-trained control filter is described in Figure \autoref{fig:Mix_Fix}. \autoref{fig:Noise_Reduction_Level_mixted} illustrates the noise reduction levels on mixed noise achieved by both methods.

The results clearly demonstrate that the FxLMS algorithm exhibits a significantly faster response to the mixed noise compared to the pre-trained control filter. During the initial 0.5 seconds, the FxLMS algorithm shows a slightly lower noise reduction performance compared to the pre-trained control filter, which can be attributed to the adaptive adjustment of the filter coefficients. In the first second, the FxLMS algorithm achieves an average noise reduction level of 13 dB, while the pre-trained control filter exhibits an average noise reduction level of 19 dB.

However, during the period from 2 to 8 seconds when the aircraft noise is particularly loud and prominent, the FxLMS algorithm surpasses the pre-trained control filter in terms of noise reduction effectiveness. Furthermore, throughout the convergence process from 11 to 13 seconds, the FxLMS algorithm consistently achieves a higher level of noise reduction compared to the pre-trained control filter. At the 13th second, the FxLMS algorithm exhibits the highest difference in average noise reduction level compared to the pre-trained control filter, with a margin of 22 dB. The FxLMS algorithm exhibits a higher signal-to-noise ratio (SNR) of 22 dB. In contrast, the pre-trained control filter achieves an SNR of 15 dB.

\begin{figure}[H]
\centering
\includegraphics[width=11.5cm]{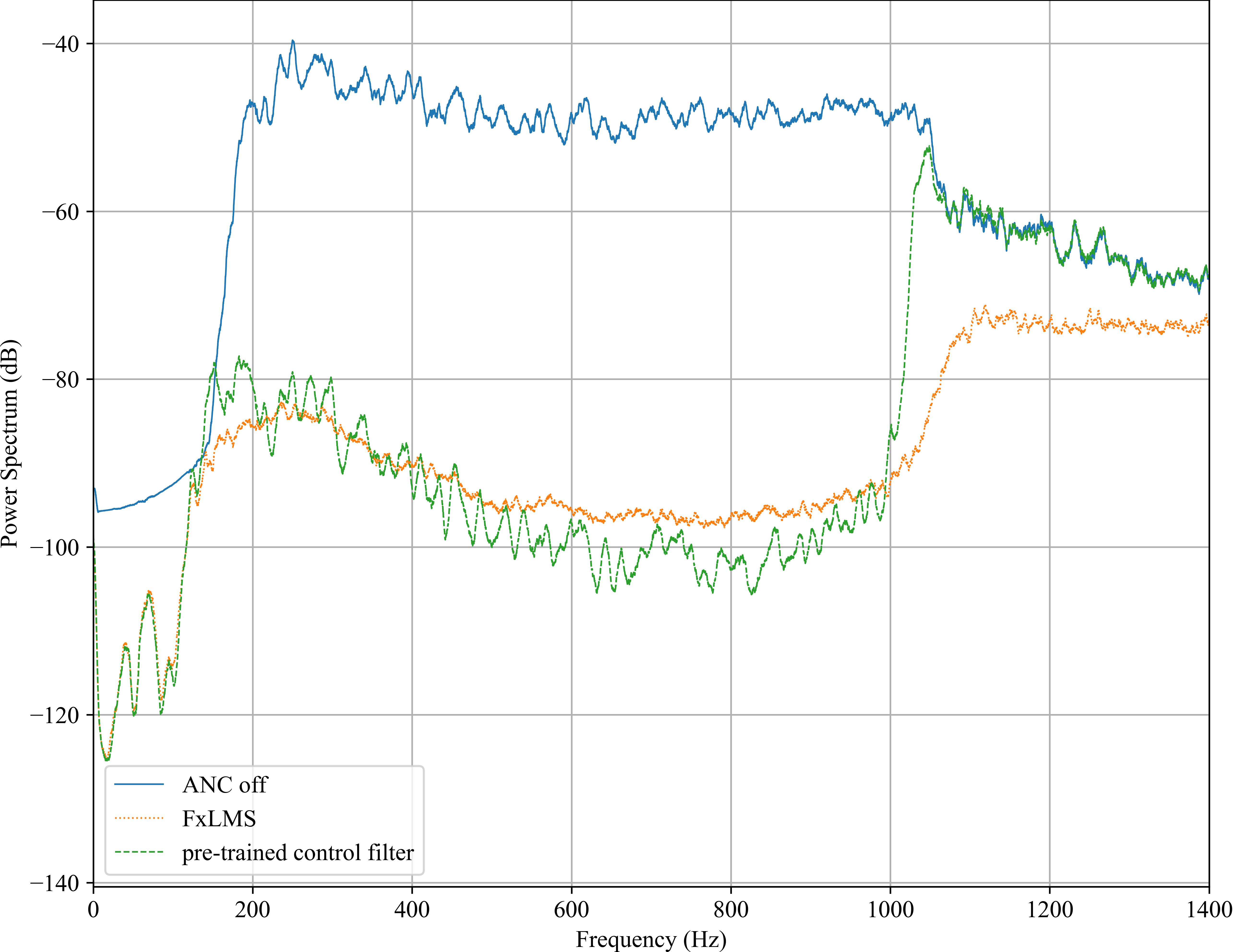}
\caption{Power spectrum of the error signal for FxLMS and pre-trained control filter in the presence of the mixed noise.}
\label{fig:Power_Spectrum_mixed} 
\end{figure}

\begin{figure}[H]
\centering
\includegraphics[width=11.5cm]{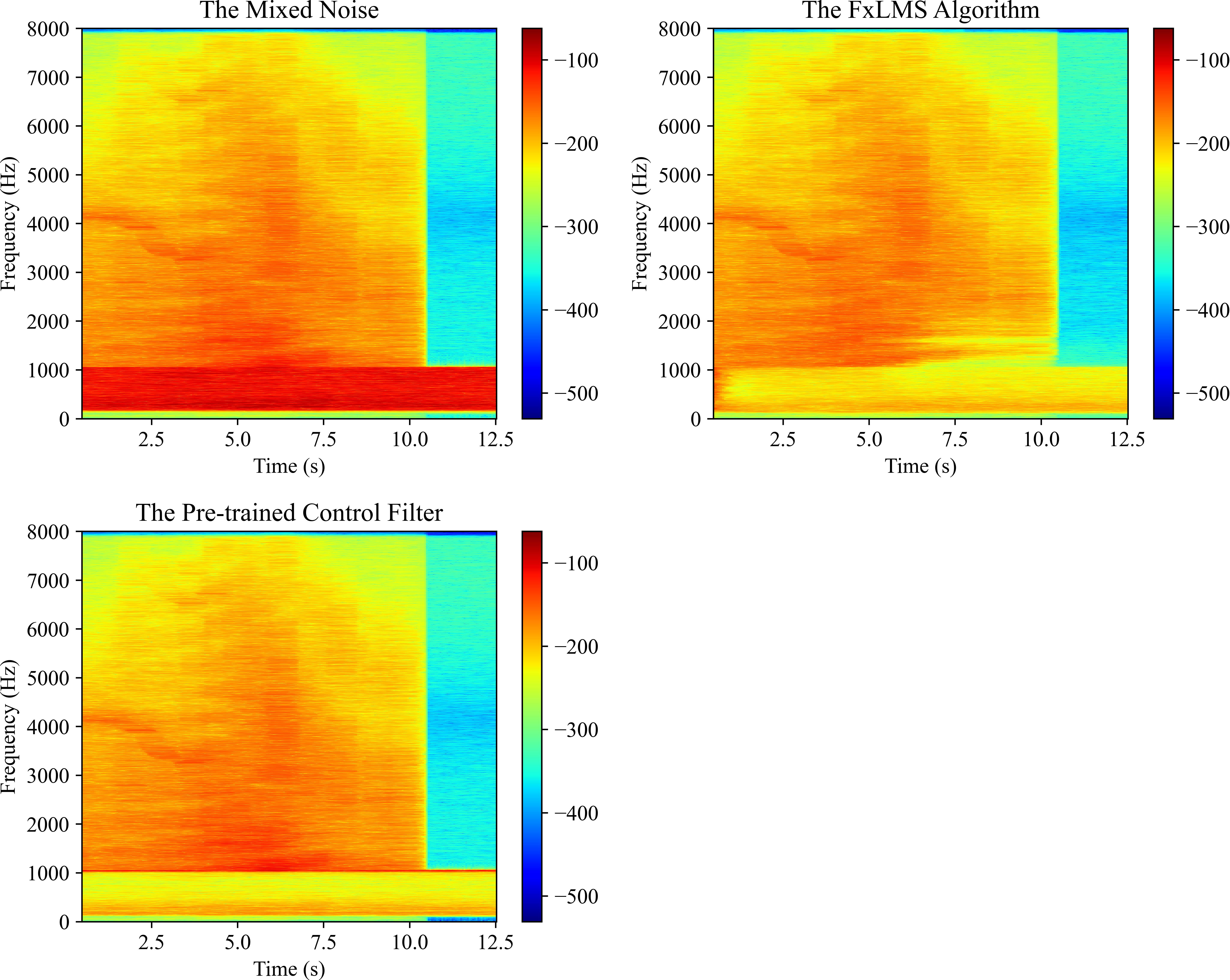}
\caption{Spectrogram of the error signal for FxLMS and pre-trained control filter in the presence of the mixed noise.}
\label{fig:Spectrogram_mixed} 
\end{figure}

For the mixed noise, when considering the overall performance across different frequency ranges, the FxLMS algorithm outperforms the pre-trained control filter in terms of noise reduction, as shown in \autoref{fig:Power_Spectrum_mixed} and \autoref{fig:Spectrogram_mixed}. The FxLMS algorithm exhibits comparable noise reduction performance to the pre-trained control filter in the low-frequency range, especially below 1000 Hz. However, in the high-frequency range, the FxLMS algorithm demonstrates better noise suppression capabilities.

From the spectrogram, it can be observed that after 0.5 seconds, the FxLMS algorithm exhibits superior noise suppression across the entire frequency range, while the pre-trained control filter only shows effectiveness in attenuating low-frequency noise.

\subsection{Discussion}
In the previous two sections, we presented and compared the noise reduction performance of the two methods in detail. We analyzed the average noise reduction levels and observed their variations over time. The results clearly demonstrated the effectiveness of both the FxLMS algorithm and the pre-trained control filter in reducing noise. Based on the presented results, several conclusions can be drawn.

The FxLMS algorithm demonstrates effective noise reduction performance on different types of real-world noise, including aircraft noise, traffic noise, and mixed noise. This indicates its versatility and suitability for various noise environments.

The pre-trained control filter, which has the advantage of pre-training the filter coefficients for a specific acoustic path, shows promising noise reduction capabilities, particularly in the initial stage of noise control.

However, as the noise control progresses, the FxLMS algorithm outperforms the pre-trained control filter in terms of average noise reduction level. This is particularly evident in the later stages of noise control, where the FxLMS algorithm consistently achieves higher levels of noise reduction.

In situations where there is a transition from one type of noise to another, such as during the combined noise scenario, the FxLMS algorithm exhibits faster response and adaptation, resulting in better noise reduction compared to the pre-trained control filter.

Overall, the FxLMS algorithm proves to be a robust and effective approach for attenuating real-world noises, showcasing its potential for practical noise control applications.

\section{Summary}

This chapter presented a comprehensive evaluation of the noise reduction performance of the FxLMS algorithm and the pre-trained control filter. Through extensive noise reduction simulations and comparisons, we gained valuable insights into the strengths and weaknesses of both algorithms in reducing real-world noise.

The comparison of the noise reduction levels, power spectrums, and spectrograms provided a detailed analysis of the performance of both methods. The findings highlighted the potential of the FxLMS algorithm for practical noise control applications. Its versatility and adaptability made it suitable for various noise environments. The results also emphasized the benefits of the pre-trained control filter in the initial stages of noise control, where it could effectively reduce noise.

\end{spacing}
\newpage


\chapter{Conclusion and Future Work}
\begin{spacing}{1.5}
\setlength{\parskip}{0.3in}
\section{Conclusion}
In this dissertation, we focused on the topic of noise control and explored the performance of two different methods, the FxLMS algorithm and the pre-trained control filter, in reducing various real-world noises. The objective was to assess their effectiveness in attenuating noise and improving the overall audio quality.

We began by providing a background on noise control and its significance in various applications. We then delved into the theoretical foundations of the FxLMS algorithm and the pre-trained control filter, explaining their underlying principles and mathematical formulations. Through this analysis, we gained a comprehensive understanding of the mechanisms behind these methods.

To evaluate the performance of the FxLMS algorithm and the pre-trained control filter, we conducted extensive simulation experiments using real-world noise signals. We specifically focused on aircraft noise, traffic noise, and mixed noise scenarios. The simulated results allowed us to compare the noise reduction capabilities of both methods and assess their effectiveness under different noise conditions.

The simulation results demonstrated that the FxLMS algorithm exhibited remarkable noise reduction performance. It showcased a superior ability to track and respond quickly to various changing noise patterns. Particularly, in the early stages of the noise reduction process, the pre-trained control filter exhibited better performance. However, as time progressed, the FxLMS algorithm consistently achieved a higher average noise reduction level compared to the pre-trained control filter. Notably, during transitional periods when noise characteristics shifted, such as from one type of noise to another, the FxLMS algorithm showcased faster responsiveness compared to the pre-trained control filter. This resilience to changes in noise frequency bands and amplitudes allowed the FxLMS algorithm to effectively attenuate noise throughout the experiment.

In summary, the experimental results demonstrated the effectiveness of the FxLMS algorithm and the pre-trained control filter in reducing real-world noises. The FxLMS algorithm, in particular, exhibited superior noise reduction capabilities, outperforming the pre-trained control filter in terms of average noise reduction levels. These findings contribute to the field of noise control and provide valuable insights for designing effective noise reduction systems.

\section{Future Work}
Future research can focus on further optimizing the FxLMS algorithm, exploring alternative adaptation strategies, and evaluating its performance in complex noise environments. Additionally, the combination of the FxLMS algorithm with other noise control techniques can be investigated to achieve even better noise reduction results. By continuously improving and refining active noise control algorithms, we can create more efficient and effective solutions for noise reduction in various settings.

\end{spacing}
\newpage


\renewcommand\bibname{References}
\bibliographystyle{unsrt}
\begin{spacing}{1.5}

\end{spacing}
\newpage

\end{document}